\documentclass[journal]{IEEEtran}

\ifCLASSINFOpdf
 \else
\fi

\usepackage{graphicx}
\usepackage{psfrag}
\usepackage{subfigure}
\usepackage{mathtools}
\usepackage{bm}
\usepackage{enumerate}
\usepackage{paralist}
\usepackage{calrsfs}
\DeclareMathAlphabet{\pazocal}{OMS}{zplm}{m}{n}
\usepackage[splitrule]{footmisc} 
\usepackage{amsfonts}
\usepackage{comment}
\usepackage[dvipsnames,table]{xcolor}
\usepackage{relsize}   
\usepackage{url}
\usepackage{footnote}

\newcommand{\dps}{\displaystyle}

\newcommand{\np}{n^\textrm{H}}
\newcommand{\npi}{n^\textrm{H,i}}
\newcommand{\npe}{n^\textrm{H,e}}

\newcommand{\tubase}{\tilde{\bm{u}}^\textrm{BC}}

\newcommand{\uopt}{\bm{u}^\textrm{OC}}
\newcommand{\tuopt}{\tilde{{\bm u}}^\textrm{OC}}
\newcommand{\tuoptprev}{\tilde{{\bm u}}^\textrm{OC,prev}}
\newcommand{\ttoptprev}{\tilde{{\bm \theta}}^\textrm{OC,prev}}
\newcommand{\ttbaseexp}{\tilde{\bm{\theta}}^\textrm{BC,exp}}
\newcommand{\ctr}{\hat{c}^\textrm{tr}}
\newcommand{\ctrDu}{\hat{c}^\textrm{tr}_{{\mathsmaller \Delta}u}}
\newcommand{\ctrDt}{\hat{c}^\textrm{tr}_{{\mathsmaller \Delta}\theta}}
\newcommand{\topt}{\bm{\theta}^\textrm{OC}}
\newcommand{\ttopt}{\tilde{\bm{\theta}}^\textrm{OC}}

\newcommand{\bx}{\bm{x}}
\newcommand{\bu}{\bm{u}}
\newcommand{\bnu}{\bm{\nu}^\textrm{m}}
\newcommand{\btheta}{\bm{\theta}}

\newcommand{\tbu}{\tilde{\bu}}
\newcommand{\tbnu}{\tilde{\bm{\nu}}}
\newcommand{\tbx}{\tilde{\bx}}
\newcommand{\bxm}{\bm{x}^\textrm{m}}
\newcommand{\bxp}{\bm{x}^\textrm{p}}
\newcommand{\bnup}{{\bm{\nu}}^\textrm{p}}

\newcommand{\BCe}{\textrm{BCe}}
\newcommand{\BCi}{\textrm{BCi}}
\newcommand{\PCe}{\textrm{PCe}}
\newcommand{\PCi}{\textrm{PCi}}
\newcommand{\nBCi}{n_{\BCi}}
\newcommand{\nBCe}{n_{\BCe}}
\newcommand{\uBCi}[1]{\bu^{\BCi(#1)}}
\newcommand{\uPCi}[1]{\bu^{\PCi(#1)}}
\newcommand{\tBCi}[1]{\btheta^{\BCi(#1)}}
\newcommand{\tPCi}[1]{\btheta^{\PCi(#1)}}
\newcommand{\uBCe}[1]{\bu^{\BCe(#1)}}
\newcommand{\nimp}{n^\textrm{i}}
\newcommand{\nexp}{n^\textrm{e}}
\newcommand{\ttBCi}[1]{\tilde{\btheta}^{\BCi(#1)}}
\newcommand{\ttPCi}[1]{\tilde{\btheta}^{\PCi(#1)}}
\newcommand{\tuBCe}[1]{\tilde{\bu}^{\BCe(#1)}}

\newcommand{\csim}{c^\textrm{s}}
\newcommand{\ks}{k^\textrm{s}}

\newtheorem{remark}{\textbf{Remark}}
\newtheorem{definition}{\textbf{Definition}}

\usepackage{etoolbox}
\makeatletter
\patchcmd{\@makecaption}
  {\scshape}
  {}
  {}
  {}
\makeatother

\makeatletter
\patchcmd{\@makecaption}
  {\\}
  {.\ }
  {}
  {}

\begin{document}

\title{Integrated Offline and Online Optimization-Based Control in a Base-Parallel Architecture}

\author{Anahita Jamshidnejad, Gabriel Gomes, Alexandre M.\ Bayen, and Bart De Schutter 
\thanks {A.\ Jamshidnejad is with the Department of Control and Simulation, 
Faculty of Aerospace Engineering, 
Delft University of Technology, Netherlands.  
G.\ Gomez and A.\ M.\ Bayen are with the Institute for Transportation Studies, 
University of California Berkeley, USA.
B.\ De Schutter is with the Delft Center for Systems and Control, 
Delft University of Technology, Netherlands.  
}}


\maketitle


\begin{abstract}
\label{sec:Abstract}
We propose an integrated control architecture to address the gap that 
currently exists for efficient real-time implementation of MPC-based 
control approaches for highly nonlinear systems with fast dynamics 
and a large number of control constraints. 
The proposed architecture contains two types of controllers: 
\emph{base controllers} that are tuned or optimized offline, and 
\emph{parallel controllers}  
that solve an optimization-based control problem online.   
The control inputs computed by the base controllers provide 
starting points for the optimization problem of the parallel controllers, 
which operate in parallel within a limited time budget that does not 
exceed the control sampling time. 
The resulting control system is very flexible and its architecture can easily 
be modified or changed online, e.g., by adding or eliminating controllers, for 
online improvement of the performance of the controlled system.
In a case study, the proposed control architecture is implemented for 
highway traffic, which is characterized by nonlinear, fast dynamics 
with multiple control constraints,  
to minimize the overall travel time of the vehicles, while increasing their total 
traveled distance within the fixed simulation time window. 
The results of the simulation show the excellent \emph{real-time} (i.e., within 
the given time budget) performance of the 
proposed control architecture, with the least realized value of the overall cost function. 
Moreover, among the online control approaches considered for the case study,   
the average cost per vehicle for the base-parallel control approach 
is the closest to the online MPC-based controllers, which have excellent performance but 
may involve computation times that exceed the given time budget.%
\end{abstract}



\IEEEpeerreviewmaketitle


\section{Introduction: objective, contributions, and structure}
\label{sec:contributions}

In this paper, we propose and develop a novel control architecture that 
integrates several control approaches in a smart and efficient way to create a
 well-performing, real-time MPC-based control approach for 
nonlinear systems with fast dynamics and a large number of control constraints. 
Our \emph{key objective} is to close the existing gap in real-time optimization-based 
control of such systems, which appear frequently in real-life applications 
and demand fast accurate control approaches.%
\smallskip
 
The main contributions of this paper include: 
\begin{itemize}
\item
We propose an integrated control system comprising multiple 
offline tuned or optimized and online optimization-based controllers 
within a novel base-parallel architecture. 
This architecture first determines several candidate control inputs for the 
controlled system at every control sampling time step, 
and next selects the one resulting in the optimal performance. 
\smallskip
\item
Our main achievement via the proposed control architecture is efficient 
\emph{real-time} (i.e., within a time budget that does not exceed the control 
sampling time) optimal control of nonlinear systems with fast dynamics 
and many control constraints, which is currently not tractable in real time.  
\smallskip
\item 
A special aspect of the proposed control architecture is its \emph{flexibility} 
for embedding and incorporating various control policies. 
Moreover, the architecture represents a partially-interconnected multi-agent 
control system with the capacity of adding or eliminating agents  
without changing the existing structure.%
\end{itemize} %
\smallskip

\noindent
The next section summarizes the main characteristics of 
offline tuned or optimized and online optimization-based controllers,   
and the existing control gaps, 
which cannot be filled with either of these control strategies alone, 
but through the integrated architecture we propose in this paper.  
In Section~\ref{sec:base_parallel_control_architecture}, 
we explain the proposed control architecture in detail. 
Section~\ref{sec:case_study} includes a case study, 
Section~\ref{sec:discussion} gives a general discussion 
of the results, and 
Section~\ref{sec:conclusions} concludes the paper and discusses 
some remaining aspects of this topic that should further be 
investigated in the future.%
 

\begin{figure*}
\centering
   \psfrag{told}[c][c][0.8]{$\btheta^\textrm{old}(\tau)$}
   \psfrag{tup}[c][c][0.8]{$\btheta^\textrm{up}(\tau)$}
   \psfrag{x}[c][c][0.8]{$\bxm(k)$}
   \psfrag{u}[c][c][0.8]{$\bu(k)$}
   \psfrag{PC}[c][c][0.8]{offline tuned controller}
   \psfrag{system}[c][c][0.8]{system}
   \psfrag{TM}[c][c][0.8]{tuning module}
   \psfrag{training data}[c][c][0.8]{\hspace*{9ex}
   $\begin{array}{l}
   \text{training dataset representing}\\
   \text{desired behaviors realized in the}\\
   \text{past and desired future behaviors}\\
   \text{predicted offline for the system}\\
   \end{array}$
   }
   \includegraphics[width = .5\linewidth]{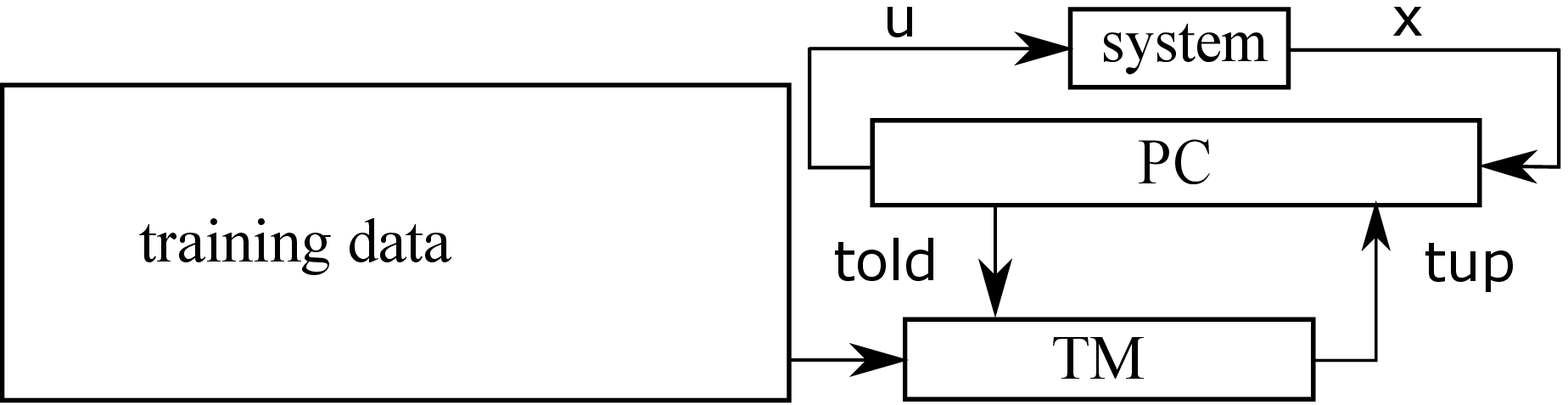}
   \caption{\emph{Offline tuned controller:} parameters are tuned initially 
   and may be updated regularly ($\tau$ is a counter for tuning sampling time steps, 
   which occur  less frequently than control sampling time steps counted by $k$).}
   \label{fig:parameterized_ctrl} 

\vspace*{2ex}

   \psfrag{offline}[c][c][0.8]{start offline}
   \psfrag{optim}[c][c][0.8]{optimization}
   \psfrag{proc}[c][c][0.8]{procedure at $k_0$}
   \psfrag{measure}[c][c][0.8]{measure system's state variables}
   \psfrag{state variable}[c][c][0.8]{state variables}
   \psfrag{predict input}[c][c][0.8]{predict external input sequence}
   \psfrag{model}[c][c][0.8]{via a prediction model}
   \psfrag{init}[c][c][0.8]{initialize $\tbu(k_0; \np)$}
   \psfrag{xm}[c][c][0.8]{$\bxm(k_0)$}
   \psfrag{tnu}[c][c][0.8]{$\tbnu(k_0 ; \np)$}
   \psfrag{tu}[c][c][0.8]{$\tbu(k_0 ; \np)$}
   \psfrag{predict}[c][c][0.8]{predict state variable sequence}
   \psfrag{compute}[c][c][0.8]{compute Bellman/value function}
   \psfrag{tx}[c][c][0.8]{$\tbx(k_0 ; \np)$}
   \psfrag{cri}[][][0.8]{criteria}
   \psfrag{sat}[][][0.8]{satisfied?}
   \psfrag{end}[][][0.8]{end}
   \psfrag{opt}[][][0.8]{optimization}
   \psfrag{proce}[][][0.8]{procedure}
   \psfrag{act}[][][0.8]{activate optimizer}
   \psfrag{adjust tu}[][][0.8]{to adjust $\tbu(k_0 ; \np)$}
   \psfrag{obj}[][][0.8]{objectives and constraints}
   \psfrag{open}[][][0.8]{{\color{red} open-loop optimization}}
   \psfrag{N}[c][c][0.8]{no}
   \psfrag{Y}[c][c][0.8]{yes}
   
   \psfrag{real}[c][c][0.8]{\hspace*{-3ex} start real-time}
   \psfrag{ctrl}[c][c][0.8]{control procedure}
   \psfrag{k=k0}[c][c][0.8]{at $ k = k_0 $}
   \psfrag{at k}[c][c][0.8]{at $ k $}
   \psfrag{model k}[c][c][0.8]{via a prediction model at $k$}
   \psfrag{init k}[c][c][0.8]{initialize $\tbu(k; \np)$}
   \psfrag{xmk}[c][c][0.8]{$\bxm(k)$}
   \psfrag{tnuk}[c][c][0.8]{$\tbnu(k ; \np)$}
   \psfrag{tuk}[c][c][0.8]{$\tbu(k ; \np)$}
   \psfrag{txk}[c][c][0.8]{$\tbx(k ; \np)$}
   \psfrag{adjust tuk}[][][0.8]{to adjust $\tbu(k ; \np)$}
   \psfrag{tuk}[c][c][0.8]{$\tbu(k ; \np)$}
   \psfrag{imp}[c][c][0.8]{implement $\bu(k)$ to controlled system,}
   \psfrag{k+1}[c][c][0.8]{implement $k \leftarrow k+1 $, and}
   \psfrag{shift forward}[c][c][0.8]{shift prediction time window one control sampling time step forward}
   
   \psfrag{closed}[][][0.8]{{\color{red} closed-loop optimization}}
   
   \includegraphics[width=\linewidth]{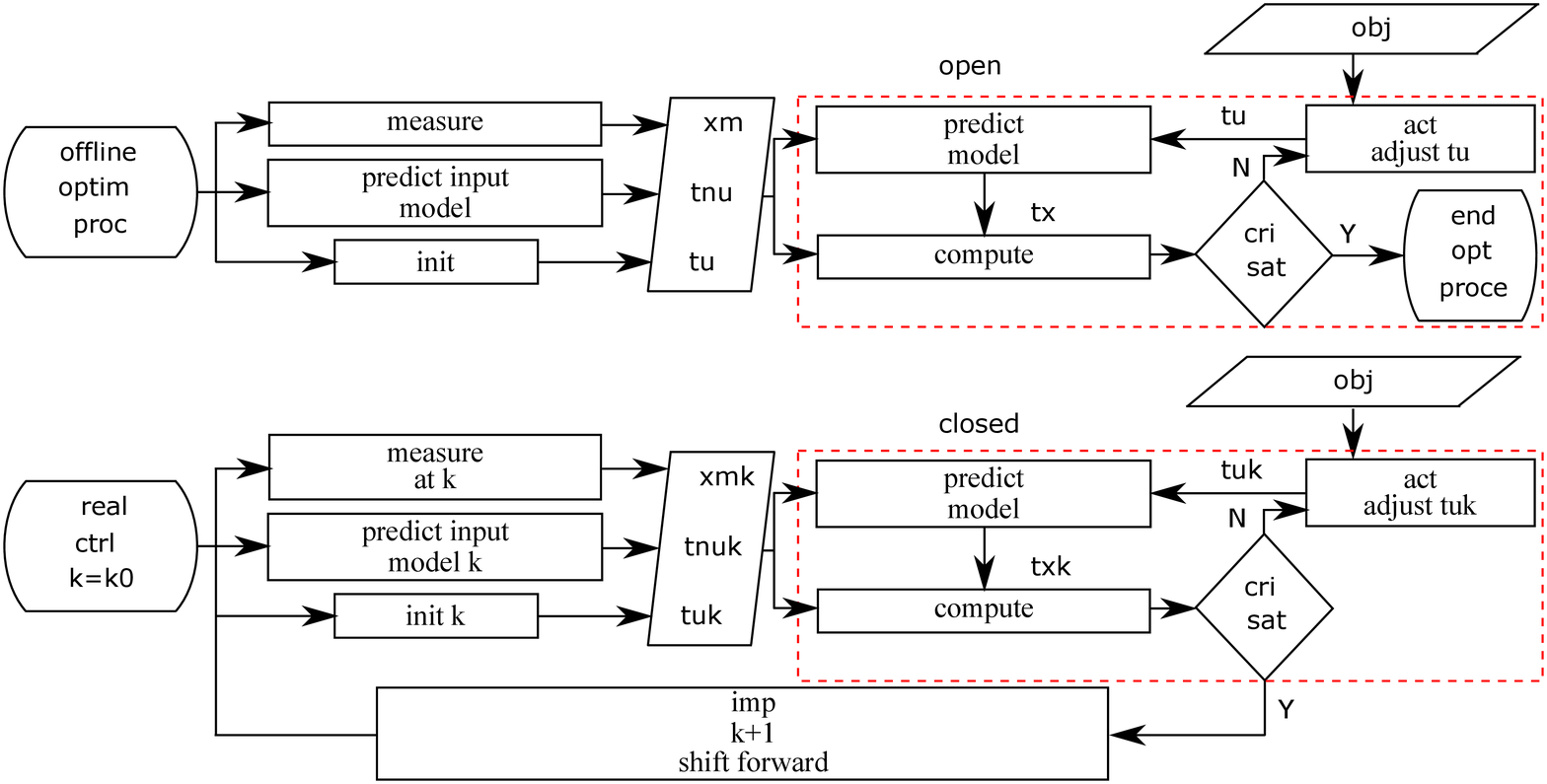}
   \caption{\emph{Optimal controller}: open-loop, i.e., classical optimal controller (top plot) 
                  and closed-loop, i.e., MPC-based controller (bottom plot), 
   with $k_0$ the initial control sampling time step and 
   $k$ the progressive control sampling time step.}
   \label{fig:optimal_ctrl}       

\end{figure*}


\section{Control categorization}
\label{sec:optimal_versus_nonoptimal_control}

Controllers may be categorized in various ways, depending on the application of interest. 
In this paper, we divide controllers into two general categories, \emph{offline} and \emph{online}, based on how the controller is tuned or optimized.%

A control law can, in general, be formulated as a mathematical expression that 
relates the control input $\bu(k)$ with the control sampling time step $k$, 
the measured state variable $\bm{x}(k)$, and the measured uncontrolled external 
input $\bnu(k)$, i.e., 
\begin{equation}
\label{eq:general_control_law}
\bu(k) = \hat{f}\left( k ,  \bxm(k) , \bnu(k) \right).     
\end{equation}
The function $\hat{f}(\cdot)$ may or may not vary while the controller runs 
(e.g., $\hat{f}(\cdot)$ may change from a polynomial to a trigonometric expression). 
Hence, an equivalent mathematical expression for \eqref{eq:general_control_law} is a parameterized formulation:
\begin{equation}
\label{eq:parameterized_control_law}
\bu(k) = \hat{u}\left( \btheta(k) , k ,  \bxm(k) , \bnu(k) \right),     
\end{equation}
where $\hat{u}(\cdot)$ has a fixed mathematical expression, but $\btheta$, i.e., the vector of the 
control input parameters, may vary.  

\smallskip
\begin{remark}
The discrete-time domain is used throughout this paper, with $k$ the 
discrete sampling time step counter.  
Bold and regular letters are used for, respectively, vectors and scalars. 
For functions, a hat symbol is used on a small regular letter. 
The superscripts ``m'' and ``p'' indicate ``measured'' and ``predicted''.
\end{remark}
\smallskip

\subsection{Offline tuned or optimized control}
\label{sec:nonoptimal_controller}

\begin{figure*}
  \centering
   \psfrag{time}[c][c][1]{time}
   \psfrag{k}[c][c][0.8]{$k$}
   \psfrag{k+1}[c][c][0.8]{$k+1$}
   \psfrag{k+2}[c][c][0.8]{$k+2$}
   \psfrag{k+3}[c][c][0.8]{$k+3$}
   \psfrag{k+4}[c][c][0.8]{$k+4$}
   \psfrag{k+5}[c][c][0.8]{$k+5$}
   \psfrag{k+6}[c][c][0.8]{$k+6$}
   \psfrag{k+7}[c][c][0.8]{$k+7$}
   \psfrag{k+8}[c][c][0.8]{$k+8$}
   \psfrag{uk}[c][c][0.8]{${\color{red}u(k)}$}
   \psfrag{uk1}[c][c][0.8]{${\color{red}u(k+1)}$}
   \psfrag{uk2}[c][c][0.8]{${\color{red}u(k+2)}$}
   \psfrag{uk3}[c][c][0.8]{${\color{red}u(k+3)}$}
   \psfrag{uk4}[c][c][0.8]{${\color{red}u(k+4)}$}
   \psfrag{uk5}[c][c][0.8]{${\color{red}u(k+5)}$}
   \psfrag{uk6}[c][c][0.8]{${\color{red}u(k+6)}$}
   \psfrag{uk7}[c][c][0.8]{${\color{red}u(k+7)}$}
   \psfrag{t1}[c][c][0.8]{${\color{red}\theta_1}$}
   \psfrag{t2}[c][c][0.8]{${\color{red}\theta_2}$}
   \psfrag{t3}[c][c][0.8]{${\color{red}\theta_3}$}
   \psfrag{ux}[c][c][0.8]{$\hat{u}\left({\color{red}\bm{\theta}} , \bm{x}(k) \right)$}
   \psfrag{ux1}[c][c][0.8]{$\hat{u}\left({\color{red}\bm{\theta}} , \bm{x}(k+1) \right)$}
   \psfrag{ux2}[c][c][0.8]{$\hat{u}\left({\color{red}\bm{\theta}} , \bm{x}(k+2) \right)$}
   \psfrag{ux3}[c][c][0.8]{$\hat{u}\left({\color{red}\bm{\theta}} , \bm{x}(k+3) \right)$}
   \psfrag{ux4}[c][c][0.8]{$\hat{u}\left({\color{red}\bm{\theta}} , \bm{x}(k+4) \right)$}
   \psfrag{ux5}[c][c][0.8]{$\hat{u}\left({\color{red}\bm{\theta}} , \bm{x}(k+5) \right)$}
   \psfrag{ux6}[c][c][0.8]{$\hat{u}\left({\color{red}\bm{\theta}} , \bm{x}(k+6) \right)$}
   \psfrag{ux7}[c][c][0.8]{$\hat{u}\left({\color{red}\bm{\theta}} , \bm{x}(k+7) \right)$}
   \psfrag{t1k}[c][c][0.8]{${\color{red}\theta_1(k)}$}
   \psfrag{t2k}[c][c][0.8]{${\color{red}\theta_2(k)}$}
   \psfrag{t3k}[c][c][0.8]{${\color{red}\theta_3(k)}$}
   \psfrag{t1k3}[c][c][0.8]{${\color{red}\theta_1(k+3)}$}
   \psfrag{t2k3}[c][c][0.8]{${\color{red}\theta_2(k+3)}$}
   \psfrag{t3k3}[c][c][0.8]{${\color{red}\theta_3(k+3)}$}
   \psfrag{t1k6}[c][c][0.8]{${\color{red}\theta_1(k+6)}$}
   \psfrag{t2k6}[c][c][0.8]{${\color{red}\theta_2(k+6)}$}
   \psfrag{t3k6}[c][c][0.8]{${\color{red}\theta_3(k+6)}$}
   \psfrag{utx}[c][c][0.8]{$\hat{u}\left({\color{red}\bm{\theta}(k)} , \bm{x}(k) \right)$}
   \psfrag{utx1}[c][c][0.8]{$\hat{u}\left({\color{red}\bm{\theta}(k)} , \bm{x}(k+1) \right)$}
   \psfrag{utx2}[c][c][0.8]{$\hat{u}\left({\color{red}\bm{\theta}(k)} , \bm{x}(k+2) \right)$}
   \psfrag{ut3x3}[c][c][0.8]{$\hat{u}\left({\color{red}\bm{\theta}(k+3)} , \bm{x}(k+3) \right)$}
   \psfrag{ut3x4}[c][c][0.8]{$\hat{u}\left({\color{red}\bm{\theta}(k+3)} , \bm{x}(k+4) \right)$}
   \psfrag{ut3x5}[c][c][0.8]{$\hat{u}\left({\color{red}\bm{\theta}(k+3)} , \bm{x}(k+5) \right)$}
   \psfrag{ut6x6}[c][c][0.8]{$\hat{u}\left({\color{red}\bm{\theta}(k+6)} , \bm{x}(k+6) \right)$}
   \psfrag{ut6x7}[c][c][0.8]{$\hat{u}\left({\color{red}\bm{\theta}(k+6)} , \bm{x}(k+7) \right)$}
   \psfrag{con}[c][c][1]{Conventional MPC}
   \psfrag{par}[c][c][1]{Parameterized MPC}
   \psfrag{B1}[c][c][1]{Move blocking MPC, Eq.~(1)}
   \psfrag{B2}[c][c][1]{Move blocking parameterized MPC, Eq.~(2)}
   \psfrag{input con}[c][c][1]{\hspace*{20ex} control input/optimization variable}
   \psfrag{input 1}[c][c][1]{\hspace*{20ex} control input/optimization variable}
   \psfrag{V}[c][c][1]{\hspace*{15ex} optimization variable}
   \psfrag{ctrl}[c][c][1]{\hspace*{10ex} control input}
   \includegraphics[width=.986\linewidth]{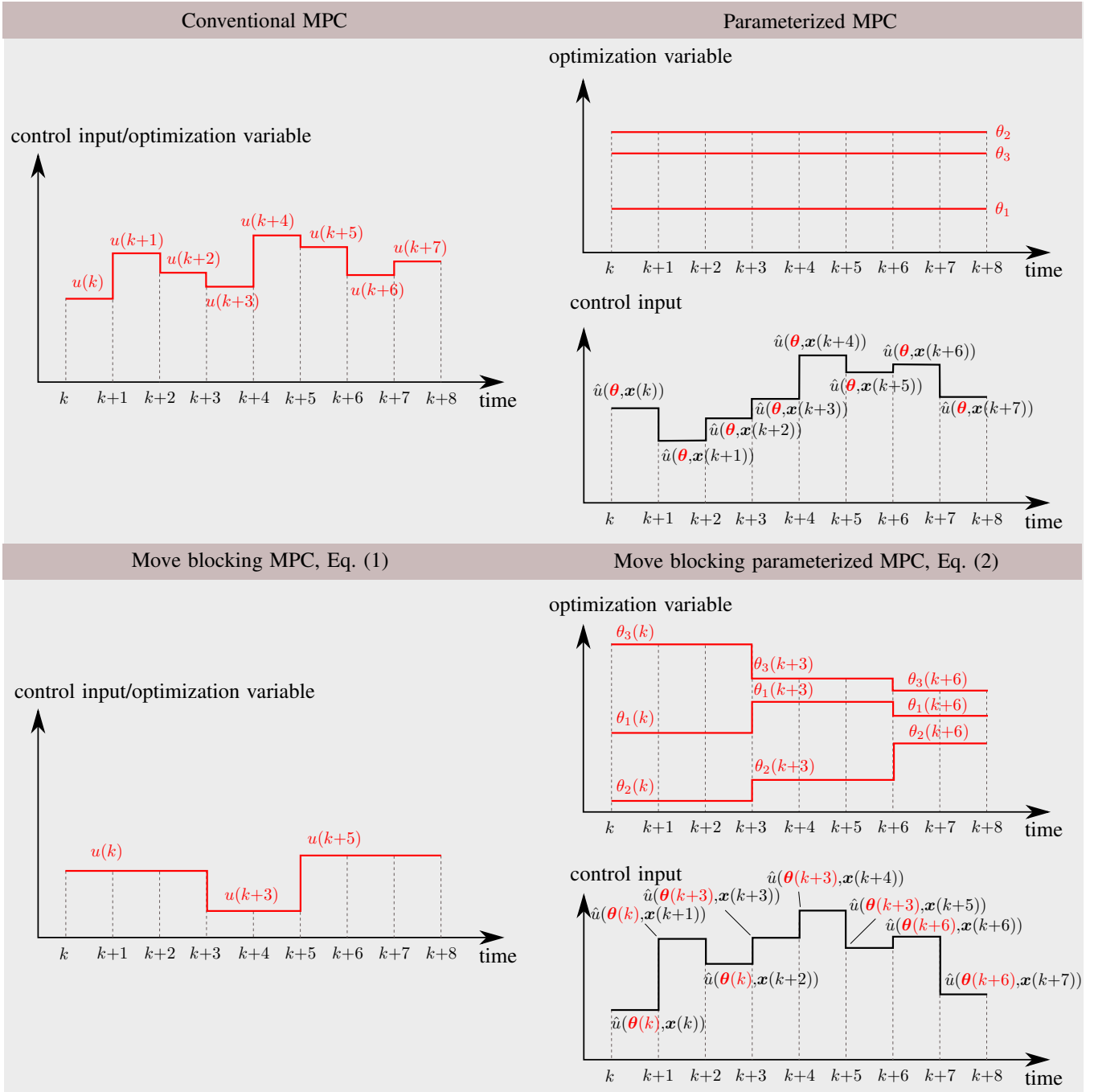}
   \caption{Conventional MPC, parameterized MPC, and move blocking MPC using 
   \eqref{eq:general_control_law} and \eqref{eq:parameterized_control_law}. 
   For the sake of compactness of the notations, we assume $\bm{\nu} = 0$ for 
   all control sampling time steps, and that $\hat{u}$ does not depend on $k$ explicitly. 
   We assume that $u$ and $\hat{u}$ are scalar. 
   The optimization variables for each case have been indicated in red.    
   We consider $\np = 8$ and a parameter vector $\btheta$ of dimension 3.}
   \label{fig:different_MPC_types}
\end{figure*}

In case the parameter vector $\btheta$ in \eqref{eq:parameterized_control_law} 
is fixed, or is tuned or optimized \emph{offline} (i.e., parallel to running the control  procedure), 
we call such control policies \emph{offline tuned or optimized} control strategies.  
To tune $\btheta$ offline, an extensive training dataset is used that 
includes pairs of state variables that cover the system's entire state space 
and control inputs that result in a desired behavior for the controlled system (see Figure~\ref{fig:parameterized_ctrl}).  
These pairs are selected in either of the following ways: 
from previous control sampling time steps, when the realized behavior of 
the controlled system has resulted in certain criteria to be satisfied, 
for instance when the performance is above a desired threshold or the CPU time has been exhausted; 
from future control sampling time steps, when the estimated behavior of the controlled 
system, using a mathematical model for predicting the state variables and 
an offline optimizer for determining the control inputs, can result in certain criteria to 
be satisfied.
%


\subsection{Online optimization-based control}
\label{sec:optimal_control}

In online optimization-based control, the parameter vector $\btheta$ 
in \eqref{eq:parameterized_control_law} is optimized 
\emph{online} at every control sampling time step, and the resulting optimal 
$\btheta$ is used in \eqref{eq:parameterized_control_law} to evaluate the 
control input $\bu(k)$ for that control sampling time step. 
The optimization is performed in a prediction time window of length $\np$ 
rather than at a single control sampling time step. 
Therefore, at every control sampling time step $k$, a sequence $\tilde{\btheta}(k;\np)$ of optimal parameter vectors, or a sequence $\tbu(k;\np)$ of the 
corresponding control inputs  are determined, computing the Bellman value function 
$b(k,\bxm(k))$ via minimizing the summation of the cumulative value of a   
performance index function $\hat{c}( \cdot,\cdot)$ within 
the prediction time window and a terminal cost value $\hat{c}^\textrm{t}\left(\bx(k + \np)\right)$, i.e., 
\begin{align}
b(k,\bxm(k))= \min_{\tbu(k;\np)} \left\{\sum_{\kappa = k}^{ k + \np - 1} 
\hat{c}(\bx(\kappa) , \bu(\kappa)) + \hat{c}^\textrm{t} (\bx(k + \np))\right\}, 
\end{align} 
subject to the physical and control constraints of the problem. 

Since the target of this paper is to develop an optimization-based 
control architecture, we next discuss some existing optimization-based 
control methods and  explain to which of the two categories introduced 
above they belong.

\begin{figure}
\centering
\psfrag{s}[][][.9]{ \hspace*{8ex} (s or higher)}
\psfrag{ms}[][][.9]{ \hspace*{8ex} (ms - $\mu$s)}
   \includegraphics[width=\linewidth]{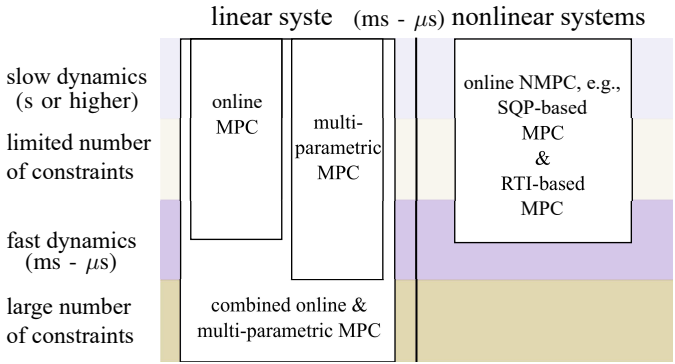}
   \caption{Applicability of various MPC-based approaches for real-time control 
   of systems with different dynamical and control characteristics. }
   \label{fig:Comparison_MPC}   
\end{figure}

\subsection{Existing optimization-based controllers}

In \emph{open-loop} optimal control (top plot in Figure~\ref{fig:optimal_ctrl}), 
the optimal control input sequence is determined offline (\emph{before} the control procedure starts)  
and the online optimization, which usually demands more time than the operational sampling time, 
is eliminated. 
Open-loop optimal control, hence, belongs to the category of offline optimized 
control. 
Since the state variables measured for solving the offline optimization problem 
may evolve considerably while the offline computations are running, 
the accuracy and reliability of the resulting control 
input sequence is under question. 
Moreover, this sequence is determined assuming perfect prediction models 
for the state variable and the external input, which in reality is not true.

In \emph{closed-loop} optimal control, i.e., conventional 
\emph{model-predictive control (MPC)} \cite{Maciejowski:2002}, 
the optimal\footnote{Note that since the length of the MPC optimization 
time window is usually smaller than the length of the control 
time window, the resulting control input sequence is, in general, suboptimal 
for the closed-loop performance.} 
control input sequence is determined online at every control sampling time step 
based on the most recent measurements of the state variable and 
the external inputs (bottom plot in Figure~\ref{fig:optimal_ctrl}). 
The first element of the optimal control input sequence is 
injected into the controlled system and the prediction 
time window is shifted forward to the next control sampling time step.  
Classically, MPC is treated as an online optimization-based control policy, 
i.e., the optimal control input sequence is determined solving the optimization 
problem of MPC online (see, e.g., \cite{Huyck:2012,Huyck:2014}). 
Online MPC should solve the optimization problem within one control 
sampling cycle, which is determined based on the speed of the dynamics of the 
controlled system and external inputs. 
This had traditionally restricted the applicability of online MPC 
to systems with slow dynamics (of order minute or second \cite{Wang:2010}). 
More recent work, e.g., \cite{Wang:2010,Houska:2011}, 
have made online linear MPC and online RTI-based nonlinear MPC   
applicable to faster dynamics (of order millisecond and microsecond).   
To tackle the computational complexities of online optimization, 
some approaches have been developed, which treat MPC as an offline 
optimized control strategy. 
Next, we discuss the following types of MPC: multi-parametric or explicit MPC 
\cite{Bemporad:2002,Johansen:2002,Bemporad:2006,Bemporad:2009}, 
combined multi-parametric and online MPC \cite{Zeilinger:2011}, 
parameterized \cite{Zegeye:2012,Muehlebach:2016} and move blocking 
\cite{Cagienard:2007,Shekhar:2015} MPC, and nonlinear MPC (NMPC) \cite{Grune:2011}.

\emph{Multi-parametric MPC} formulates the MPC control inputs as a set of 
explicit functions of the system's state variables. 
Each function is an \emph{offline} solution of the optimization problem 
in a particular subregion of the system's state space, treating the 
initial state as a parameter. The online computations are 
reduced to the evaluation of the corresponding function for the measured state.   
Hence, multi-parametric MPC lies within the category of offline optimized control. 
In general, the number of the state space subregions grows exponentially with 
the number of the state components and constraints, which implies a large CPU 
time for the computations and huge memory requirements for storing the parametric solutions. 
Bemporad and Filippi \cite{Bemporad:2003} propose approximate solutions
(by relaxing some of the first-order KKT optimality conditions)
 for explicit MPC with a reduced number of state space subregions, which     
despite the positive influence on the CPU time and required memory, 
may however negatively affect the optimality.

The next MPC strategies belong to the online optimization-based control 
category.  
In combined multi-parametric and online MPC \cite{Zeilinger:2011}, 
the control input sequence evaluated by multi-parametric MPC is used at 
every control sampling time step as a starting optimization point for 
the online MPC. 
This method has been applied to quadratic programming in \cite{Zeilinger:2011}.

In contrast to conventional MPC (see the top left plot in Figure~\ref{fig:different_MPC_types}), 
where $\hat{f}\left(\cdot\right)$ in 
\eqref{eq:general_control_law} or $\btheta$ in \eqref{eq:parameterized_control_law} 
are optimized for  every control sampling time step in the prediction time  window, 
in \emph{parameterized MPC} and \emph{move blocking MPC}, which lie within the category 
of online MPC, $\hat{f}\left(\cdot\right)$ or $\btheta$ 
are optimized subject to an extra constraint that reduces the degree of freedom. 
In parameterized MPC, $\btheta$ is fixed for all control sampling time steps within the 
prediction time window, and the number of the elements of the parameter vector $\btheta$ 
is considered less than the number of elements of the original control input sequence   
(see the top right plot in Figure~\ref{fig:different_MPC_types}). 
In move blocking MPC, $\hat{f}\left(\cdot\right)$ or $\btheta$ follow a blocking 
stepwise function, i.e., the optimization variables are constant for multiple consecutive control 
sampling time steps (see the bottom plots in Figure~\ref{fig:different_MPC_types}).  
In summary, in parameterized and move blocking MPC the number of the optimization 
variables compared to conventional MPC is reduced, so that the computation 
time may become smaller. 
A main drawback is that since the solutions are limited to the family 
of the introduced parametric 
function, only a subregion of the feasible optimization region is searched, 
while the optimal value(s) of the original optimization problem may or 
may not belong to this subregion.%

NMPC is preferred over linear MPC when 
the nonlinear dynamics and constraints are better treated explicitly 
(without linearization) due to the accuracy requirements. 
Initially, online NMPC was used in the process industry for nonlinear systems 
with slow dynamics \cite{Grune:2011,Chen:2000}. 
Nowadays, there are fast NMPC approaches, such as the 
quadratic programming (SQP) and real-time iteration (RTI)
that can operate online for control sampling times in the range of millisecond 
\cite{Houska:2011,Alamir:2014,Gros:2016}.%


\subsection{Existing gaps in online MPC-based control}

Figure~\ref{fig:Comparison_MPC} illustrates, in one glance, the fields of 
online applicability of the discussed MPC-based approaches. 
For nonlinear systems  involving fast dynamics and a large number 
of constraints, a gap exists that cannot yet be 
addressed via the existing MPC-based control approaches.   
Despite the good performance of NMPC and RTI-based MPC with dynamics as 
fast as millisecond and microsecond \cite{Alamir:2014,Houska:2011}, 
as the complexity level, e.g., nonlinearity, and the  number of constraints rise, 
the applicability of these approaches becomes more restricted.%

The theory and application of online MPC for nonlinear systems 
needs to be further developed. \emph{The main aim of this paper} is  
to take significant steps towards closing this gap by proposing a novel control 
architecture that integrates several control approaches in a smart and efficient way  
to create a well-performing, real-time\footnote{The definition of ``real-time'' in this 
paper is relative, i.e., it depends on the dynamics evolution of the system and 
correspondingly, the control sampling time. A control approach is real-time for a 
specific controlled system, if the computation time of the control approach 
does not exceed the control sampling time of the controlled system.} MPC-based control approach for 
nonlinear systems with a large number of constraints.%


\section{Base-Parallel Integrated Control Architecture}
\label{sec:base_parallel_control_architecture}

\begin{figure}\centering
\includegraphics[width = .5 \linewidth]{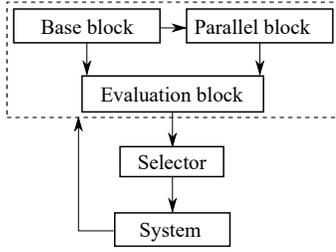}
\caption{Base-parallel integrated control architecture for efficient real-time control of nonlinear systems with fast dynamics and many constraints.}
\label{fig:simplified_base_parallel_block}
\end{figure}

\begin{figure*}
\psfrag{base block}[][][.75]{\textbf{Base block}}
\psfrag{parallel block}[][][.75]{\textbf{Parallel block}}
\psfrag{Evaluation block}[][][.75]{\textbf{Evaluation block}}
\psfrag{BCi1}[][][.75]{$\BCi(1)$}
\psfrag{PCi11}[][][.75]{$\PCi(1,1)$}
\psfrag{PCi1n}[][][.75]{$\PCi(1,\nimp(1))$}
\psfrag{BCin}[][][.75]{$\BCi(\nBCi)$}
\psfrag{PCin1}[][][.75]{$\PCi\big(\nBCi,1\big)$}
\psfrag{PCinn}[][][.75]{$\PCi\big(\nBCi,\nimp(\nBCi)\big)$}
\psfrag{eval 2}[][][.75]{Evaluate \eqref{eq:parameterized_control_law}}
\psfrag{ubi1}[][][.75]{$\uBCi{1}(\kappa^\textrm{i}(i_1))$}
\psfrag{ubin}[][][.75]{$\uBCi{\nBCi}(\kappa^\textrm{i}(i_{\nBCi})$}
\psfrag{thetab1}[][][.75]{$\tBCi{1}(\kappa^\textrm{i}(i_1))$}
\psfrag{thetabn}[][][.75]{$\tBCi{\nBCi}\left(\kappa^\textrm{i}(i_{\nBCi}) \right)$}
\psfrag{tthetab1}[][][.75]{$\ttBCi{1}(k;\npi(1,i_1))$}
\psfrag{tthetabn}[][][.75]{$\ttBCi{\nBCi}(k;\npi(\nBCi,i_{\nBCi}))$}
\psfrag{nb1}[][][.75]{$\max\left\{\npi(1,i_1)\right\}$}
\psfrag{nbn}[][][.75]{$\max \left\{ \npi(\nBCi,i_{\nBCi}) \right\}$}
\psfrag{Mi}[][][.75]{\textrm{M}}
\psfrag{x1}[][][.75]{$\bxp(\kappa^\textrm{i}(i_1) + 1),$}
\psfrag{v1}[][][.75]{$\bnup (\kappa^\textrm{i}(i_1) + 1)$}
\psfrag{xn}[][][.75]{$\bxp(\kappa^\textrm{i}(i_{\nBCi}) + 1),$}
\psfrag{vn}[][][.75]{$\bnup (\kappa^\textrm{i}(i_{\nBCi}) + 1)$}
\psfrag{ube1}[][][.75]{$\uBCe{1}(\kappa^\textrm{e}(e_1))$}
\psfrag{uben}[][][.75]{$\uBCe{\nBCe}(\kappa^\textrm{e}(e_{\nBCe}))$}
\psfrag{tube1}[][][.75]{$\tuBCe{1}(k;\npe(1,e_1))$}
\psfrag{tuben}[][][.75]{$\tuBCe{1}(k;\npe(\nBCe,e_{\nBCe}))$}
\psfrag{nbe1}[][][.75]{$\max \left\{\npe(1,e_1) \right\}$}
\psfrag{nben}[][][.75]{$\max\left\{ \npe(\nBCe,e_{\nBCe}) \right\}$}
\psfrag{xe1}[][][.75]{$\bxp(\kappa^\textrm{e}(e_1) + 1),$}
\psfrag{ve1}[][][.75]{$\bnup (\kappa^\textrm{e}(e_1) + 1)$}
\psfrag{xen}[][][.75]{$\bxp(\kappa^\textrm{e}(e_{\nBCe}) + 1),$}
\psfrag{ven}[][][.75]{$\bnup (\kappa^\textrm{e}(e_{\nBCe}) + 1)$}
\psfrag{E}[][][.75]{E}
\psfrag{tthetap1}[][][.75]{$\ttPCi{1,i_1}(k;\npi(1,i_1))$}
\psfrag{thetap1}[][][.75]{$\tPCi{1,i_1}(k)$}
\psfrag{upi1}[][][.75]{$\uPCi{1,i_1}(k)$}
\psfrag{thetapn}[][][.75]{$\tPCi{\nBCi,i_{\nBCi}}(k)$}
\psfrag{upin}[][][.75]{$\uPCi{\nBCi,i_{\nBCi}}(k)$}
\psfrag{uBC}[][][.75]{$\bu^{\textrm{BC}}(k)$}
\psfrag{uPC}[][][.75]{$\bu^{\textrm{PC}}(k)$}
\psfrag{c(k)}[][][.75]{$\bm{\epsilon} (k)$}
\psfrag{xm}[][][.75]{$\bm{x}^\textrm{m}(k) , \bm{\nu}^\textrm{m}(k)$}
\psfrag{BCe1}[][][.75]{$\BCe(1)$}
\psfrag{PCe11}[][][.75]{$\PCe(1,1)$}
\psfrag{PCe1n}[][][.75]{$\PCe(1,\nexp(1))$}
\psfrag{BCen}[][][.75]{$\BCe(\nBCe)$}
\psfrag{PCen1}[][][.75]{$\PCe\big(\nBCe,1\big)$}
\psfrag{PCenn}[][][.75]{$\PCe\big(\nBCe,\nexp(\nBCe)\big)$}
\psfrag{ube1}[][][.75]{$\uBCe{1}$}
\psfrag{uben}[][][.75]{$\uBCe{\nBCe}$}
\psfrag{ubest(k)}[][][.75]{$\bu^\textrm{best} (k)$}
\psfrag{BCi}[][][.75]{BCi: implicit base controller}
\psfrag{BCe}[][][.75]{BCe: explicit base controller}
\psfrag{PCi}[][][.75]{PCi: implicit parallel controller}
\psfrag{PCe}[][][.75]{PCe: explicit parallel controller}
\psfrag{Mdef}[][][.75]{\hspace*{25ex} M: mathematical model for the state and uncontrolled external inputs}
\psfrag{Edef}[][][.75]{E: evaluator}
\psfrag{nHi}[][][.75]{$\npi$: prediction horizon of an implicit parallel controller}
\psfrag{nHe}[][][.75]{$\npe$: prediction horizon of an explicit parallel controller}
\psfrag{nBCi}[][][.75]{$\nBCi$: total number of implicit base controllers within the base block}
\psfrag{nBCe}[][][.75]{$\nBCe$: total number of explicit base controllers within the base block}
\psfrag{ni}[][][.75]{$\nimp$: total number of implicit parallel controllers within a 
  specific implicit parallel cell}
\psfrag{ne}[][][.75]{$\nexp$: total number of explicit parallel controllers within a 
  specific explicit parallel cell}
\psfrag{e}[][][.75]{$\bm{\epsilon}(k)$: vector including the realized values of the performance indices}
\psfrag{e1}[][][.75]{for all the candidate control inputs at control sampling time step  $k$}
\includegraphics[width=\linewidth]{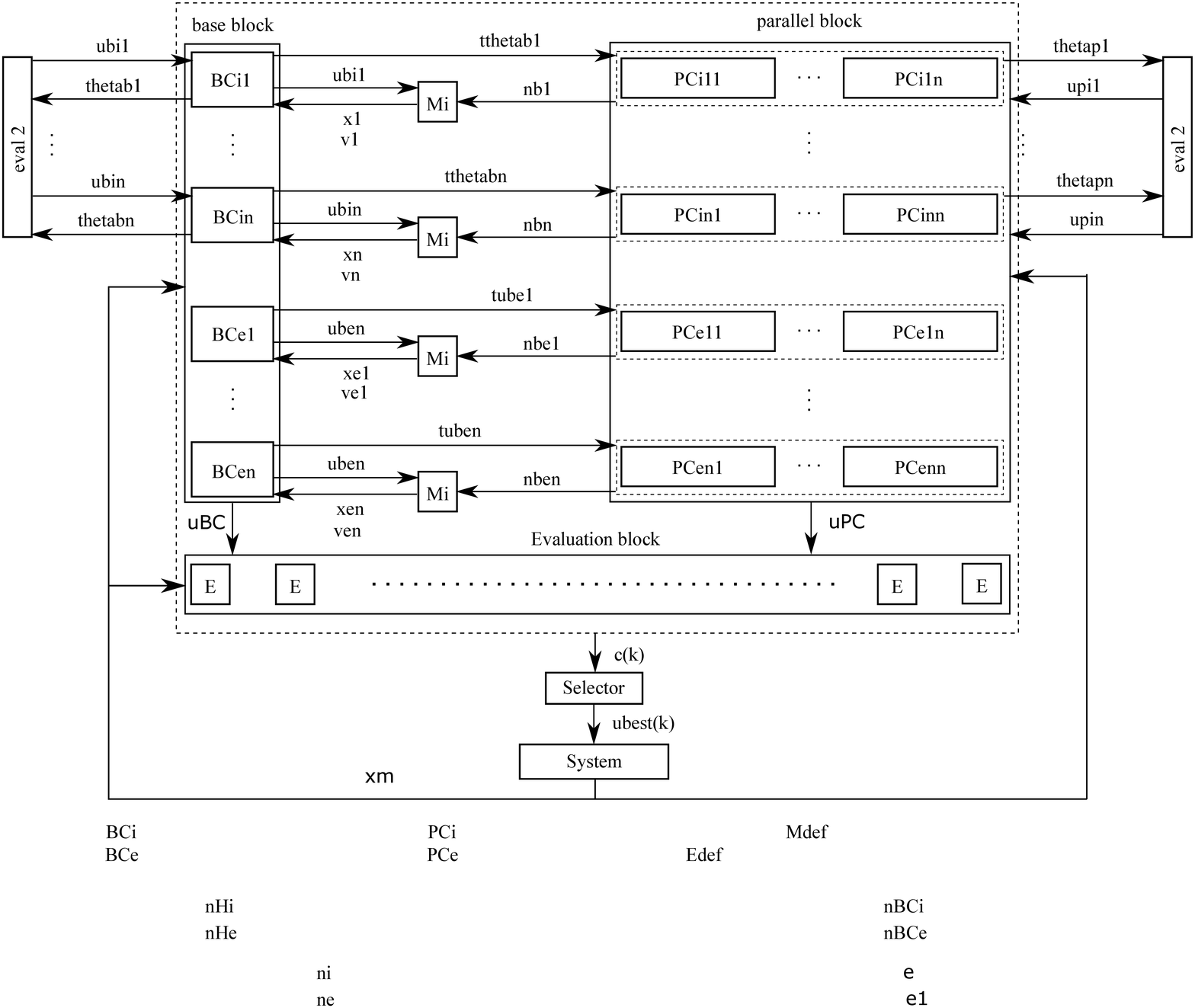}
\centering
\caption{Detailed illustration of the base-parallel integrated control architecture.}
\label{fig:Base_parallel_control_architecture}
\end{figure*}


In this section, the proposed integrated control architecture 
is explained in detail.  
Figure~\ref{fig:simplified_base_parallel_block} shows the main components 
of the control architecture, i.e., the \emph{base block} 
and the \emph{parallel block}. 
The base block consists of multiple controllers that are tuned or optimized 
offline, called \emph{base controllers}. 
The parallel block includes several online optimization-based controllers, called \emph{parallel 
controllers}, which run in parallel during the \emph{online} control procedure. 
The control architecture, also including an \emph{evaluation block} composed of 
identical mathematical models of the controlled system and a \emph{selector}, is feedback-based.
Some measurements, e.g., the realized state variables, the uncontrollable external inputs, 
and the performance indices, from the controlled system are fed back into the control architecture. 
The different control modules, which have been illustrated in detail in 
Figure~\ref{fig:Base_parallel_control_architecture}, will next be explained.%


\subsection{Base control block}
\label{sec:base_control_block}

Considering the \emph{structure of the control policy}, 
%
a base controller may be \emph{fixed-time}, i.e., the function $\hat{f}(\cdot)$ 
in \eqref{eq:general_control_law} or equivalently the parameter vector $\btheta$ 
in \eqref{eq:parameterized_control_law} are fixed in time. 
Therefore, the same mathematical formulation 
is used at all control sampling time steps to evaluate the control input $\bu$  
from the measured values of the state variable and uncontrolled external inputs. 
Control policies such as  open-loop optimal control, multi-parametric MPC, 
full-state feedback control (pole placement), offline optimized 
state feedback control, and PID control may be optimized or tuned offline and used 
as fixed-time base controllers.%

\begin{figure}
\psfrag{base}[][][.8]{Trained mapping (e.g., ANN):}
\psfrag{t=m(x,v)}[][][.8]{$\btheta(k) = \hat{\mu}\left(\bxm(k),\bnu(k)\right)$}
\psfrag{t}[][][.8]{computation time for mapping evaluation $\approx$ 0}
\psfrag{x}[][][.8]{$\bxm(k)$,$\bnu(k)$}
\psfrag{theta}[][][.8]{$\btheta(k)$}
\psfrag{optim}[][][.8]{Online optimizer with a recursive procedure:}
\psfrag{t>0}[][][.8]{computation time for optimization $\gg$ 0} 
\psfrag{min}[][][0.8]{
$\min\limits_{\tilde{\btheta}(k;\np)} \left\{
\sum\limits_{\kappa = k}^{ k + \np - 1} 
\hat{c}(\bx(\kappa) , \btheta(\kappa)) + 
\hat{c}^\textrm{t} \left(\bx \left(k + \np \right) \right)
\right\}$}
\psfrag{const}[][][0.8]{subject to constraints}
\psfrag{input}[][][0.7]{input}
\psfrag{output}[][][0.7]{output}
\psfrag{est}[][][0.8]{estimate future}
\psfrag{state}[][][0.8]{states}
\psfrag{opt}[][][0.8]{optimize parameter}
\psfrag{v}[][][0.8]{vector}
\includegraphics[width = \linewidth]{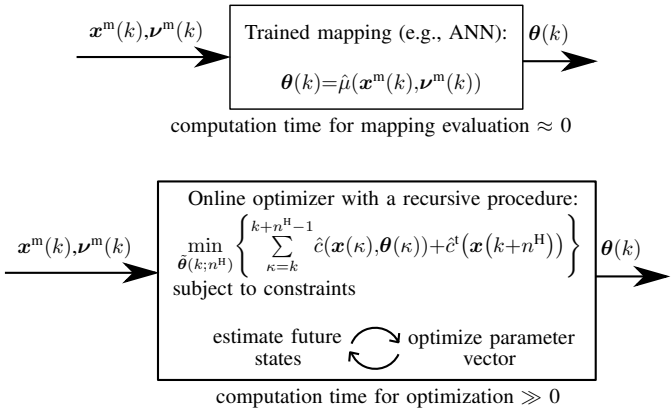}
\caption{Time-varying base controller vs.\ online optimization-based controller, 
         where $\hat{\mu}(\cdot)$ is a trained mapping.}
\label{fig:compare_ctrl}
\end{figure}


Some base controllers may be \emph{time-varying}, i.e.,  the function 
$\hat{f}(\cdot)$ in \eqref{eq:general_control_law} or equivalently the 
parameter vector $\btheta$ in \eqref{eq:parameterized_control_law}  
can vary in time. 
A mapping, e.g., an artificial neural network (ANN), 
that has been formulated and trained offline using an extensive dataset 
collected via performing simulations or from the past desired behavior of 
the controlled system, is used to evaluate $\btheta$ or equivalently the structure 
of $\hat{f}(\cdot)$. This evaluation may be done at every control sampling time step or 
less frequently.  
Note that since the trained mapping involves algebraic computations, the online 
computation time is negligible. 
This is in contrast to an optimizer that involves an online optimization 
procedure with recursive differential computations and estimations, 
and hence may suffer from online computational complexities (see 
Figure~\ref{fig:compare_ctrl}).  

\smallskip

Considering the \emph{structure of the output} of a base controller, the 
following two categories may exist: 
\subsubsection{Explicit base controller}
Such a controller produces the control input $\bu$ directly 
at every control sampling time step. 
\subsubsection{Implicit base controller}
Such a controller produces the parameter vector $\btheta$ in 
\eqref{eq:parameterized_control_law} at every control sampling time step. 


\subsection{Parallel control block}
\label{sec:parallel_control_block}

Considering the \emph{structure of the output}, a parallel controller 
may belong to either of the following two categories:

\subsubsection{Explicit parallel controller}

At every control sampling time step $k$, such a controller directly 
optimizes the  sequence $\tilde{\bu}\left( k ; \np \right)$ 
of the control inputs for all the control sampling time steps within the prediction time window. 
A conventional MPC-based controller can, e.g., be used as an explicit parallel controller.

\subsubsection{Implicit parallel controller}

At every control sampling time step $k$, such a controller uses the formulation \eqref{eq:parameterized_control_law} 
to optimize the sequence $\tilde{\btheta}(k ; \np)$ of the control parameter vectors 
within the prediction time window. 
A  parameterized MPC-based controller is an example of an implicit parallel controller.%

\smallskip

A CPU time budget is given to the parallel block for solving the  
optimization problems. When this time budget is exhausted, 
all the optimization procedures will be terminated. 
If the optimization problem has been solved within the given time budget, 
the optimal solution will be the candidate control input sequence of the corresponding parallel controller. 
Otherwise, if an optimization procedure has been terminated before an 
optimal solution that satisfies the optimization criteria was found, 
two options are possible:
\begin{compactenum}
\item
The realized values of the cost for all optimization iterations are saved together with 
their corresponding control input sequences. After the termination of the optimization, 
these values are compared and the control input sequence that 
corresponds to the least realized optimization cost is selected 
as the candidate control input of that parallel controller.  
\item 
All control input sequences that were determined at the optimization iterations are 
injected into the evaluation block as candidate control input sequences of that 
parallel controller. Note that this option can be beneficial when the mathematical 
model used in the evaluation block (see Section~\ref{sec:evaluation_block_and_selector} 
for more details) is different from (i.e., contains more details than) the prediction model 
of the parallel controller, and the computational burden for evaluating 
all the cost values corresponding to these candidate control input sequences  
by the evaluation block is affordable for the control system.  
\end{compactenum}


\subsection{Integrated structure of the base and parallel blocks}
\label{sec:integration}

Figure~\ref{fig:Base_parallel_control_architecture} illustrates the proposed 
integrated control architecture in more detail. 
At every control sampling time step $k$, the measured values 
of the state variable, $\bm{x}^\textrm{m}(k)$, and uncontrolled external input, 
$\bm{\nu}^\textrm{m}(k)$, are injected into the base, parallel, and evaluation blocks. 
The implicit and explicit base controllers, indicated by $\BCi$ and 
$\BCe$ in the figure, will evaluate the control parameter vector 
$\btheta^{\BCi}(k)$ and the control input $\bu^{\BCe}(k)$ respectively, 
using these measurements. 
The vectors $\btheta^{\BCi}(k)$ are used to evaluate the corresponding 
control inputs $\bu^{\BCi}(k)$ using \eqref{eq:parameterized_control_law}. 
The  vector $\bu^\textrm{BC}(k)$, including  all candidate control inputs 
$\bu^{\BCe}(k)$ and $\bu^{\BCi}(k)$ from the explicit and implicit base controllers,    
is injected into the evaluation block.%

Moreover, in order for the parallel controllers to start from a starting optimization 
point with a higher chance of converging to an optimum within the given CPU 
time budget, the control inputs corresponding to the base controllers are injected 
into the parallel block as starting points for the optimization. 
This is a key aspect of the proposed integrated control architecture. 
Before we explain the details of how this idea is implemented in the integrated 
base-parallel architecture, the terminology that will be used is explained. 
In Figure~\ref{fig:Base_parallel_control_architecture}, every row of  
parallel controllers that is indicated by a dashed rectangle is 
called a \emph{parallel cell}. 
The elements of the optimization variable sequences of the parallel controllers 
in one parallel cell are of the same nature, i.e., either explicit control input vectors 
or control input parameter vectors.  
However, the number of the elements in the sequences, i.e., the control horizon of 
the parallel controllers in one cell, may be different from each other. 
Therefore, a parallel cell is either  an \emph{explicit parallel cell}, i.e., 
all parallel controllers in the cell are explicit, or is an \emph{implicit parallel cell}, 
i.e., all parallel controllers in the cell are implicit. 
Each parallel cell corresponds to a particular base controller, 
i.e., the parallel controllers in that parallel cell receive the starting 
optimization sequence computed by this particular base controller.  
Hence, an explicit parallel cell corresponds to an explicit base controller, 
and an implicit parallel cell corresponds to an implicit base controller.%

\begin{figure*}
\centering
\psfrag{base}[][][.85]{base controller}
\psfrag{par}[][][.85]{parallel controller}
\psfrag{est}[][][.85]{estimate $\bu(k+1),\ldots,\bu(k+\np-1)$}
\psfrag{u}[][][.85]{$\bu(k)$}
\psfrag{seq}[][][.85]{$\left\{ \bu(k+1),\ldots,\bu(k+\np-1)\right\}$}
\psfrag{u0}[][][.85]{starting optimization point}
\includegraphics[width = .85\linewidth]{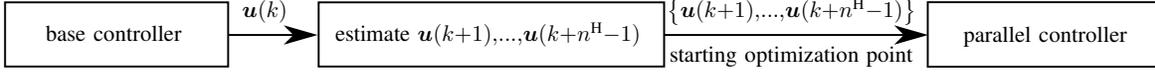}
\caption{The control input $\bu(k)$ produced by the base controller should 
first be used to estimate the control input for the future sampling control time 
steps in the prediction time window. This estimation is done by a mathematical   
model of the controlled system and the uncontrolled external inputs.}
\label{fig:base_estimate_start_optimization}
\end{figure*}

Since the parallel controllers follow model-based and optimization-based control 
strategies  that compute  the control input or the control parameter vector within 
a prediction time window (including the current and the $\np - 1$ future control 
sampling time steps), these controllers optimize a \emph{sequence} of 
either the control input or the control parameter vector.   
Therefore, the parallel controllers should initially receive a sequence of the 
control input or control parameter vector to start the optimization procedure. 
The base controllers, however, produce the control input or control parameter 
vector for the current control sampling time step only. Hence, the control input or 
the control parameter  vector should first be estimated for the control sampling   
time steps $k+1, \ldots, k+\np -1$ based on the one computed by the base controller 
for the current control sampling time step (see Figure~\ref{fig:base_estimate_start_optimization}).%

A mathematical model of the states of the controlled system, integrated with a mathematical model 
of the uncontrolled external inputs is used for this estimation. 
This model should perform in a loop (of size the largest prediction horizon of 
the corresponding parallel controllers) together with the base controller: 
for the control sampling time steps $\kappa \in \left\{ k,\ldots,k+\np-1 \right\}$, 
the control input at the control sampling time step $\kappa$ computed by the base controller 
is sent to the integrated mathematical model, which estimates the states of the controlled 
system based on the received control input, and the uncontrolled external inputs 
for the next control sampling time step $\kappa+1$. 
These estimated values are sent back to the base controller to compute the control input at $\kappa+1$. 
This loop has been illustrated in Figure~\ref{fig:Base_parallel_control_architecture}.  
The size of the loop is determined via the size of the prediction horizon of the 
corresponding parallel controllers. Therefore, the maximum size of all the prediction horizons 
in the parallel cell (see Figure~\ref{fig:Base_parallel_control_architecture}) is given to the 
loop, such that the loop generates a sequence of control inputs of this maximum size. 
Every parallel controller in the parallel cell receives as the starting point for the optimization  
the first element of this sequence to the element of the number of its prediction horizon.   
Note that for an implicit base controller, the control input is first evaluated via 
\eqref{eq:parameterized_control_law} using the control parameter vector produced by the implicit base 
controller.%

\subsection{Evaluation block and selector}
\label{sec:evaluation_block_and_selector}

At every control sampling time step $k$, all the control inputs determined by the base and parallel controllers (indicated by $\bu^\textrm{BC}(k)$ and $\bu^\textrm{PC}(k)$ 
in Figure~\ref{fig:Base_parallel_control_architecture})  enter the 
evaluation block. This block consists of identical integrated mathematical models,  
which estimate the states of the controlled system based on the received control 
input, and also estimate the uncontrolled external inputs. 
Note that since the evaluation block has to run only once for every control sampling time step, 
these integrated models can be much more refined and detailed than the 
models used for the base and parallel controllers.  
The number of these 
models, in general, equals the total number of the base and parallel 
controllers in the integrated control architecture.  
These models will run in parallel to evaluate a predefined performance   
index function  $\hat{\epsilon}(\cdot)$ for the controlled system for each received 
candidate control input. The resulting values (indicated by vector $\bm{\epsilon}(k)$ 
in Figure~\ref{fig:Base_parallel_control_architecture}) are sent to a selector 
that compares them and selects the control input corresponding to   the  
least realized value of the performance index as the best candidate $\bu^\textrm{best}(k)$ 
at control sampling time step $k$ to be injected into the controlled system.%

\section{Case Study}
\label{sec:case_study} 

In this section, we present the results of a case study, where the proposed 
integrated base-parallel control architecture is implemented to a highway stretch  
to increase the total distance traveled by the vehicles during the fixed 
simulation time window, while reducing their total travel time.%

\subsection{Highway simulation}
\label{sec:highway_simulation}

\begin{figure}
\centering
\psfrag{i-1}[][][.75]{cell $i-1$}
\psfrag{i}[][][.75]{cell $i$}
\psfrag{i+1}[][][.75]{cell $i+1$}
\psfrag{f1}[][][.75]{$o_{i-2}(\ks)$}
\psfrag{f2}[][][.75]{$o_{i-1}(\ks)$}
\psfrag{f3}[][][.75]{$o_{i}(\ks)$}
\psfrag{f4}[][][.75]{$o_{i+1}(\ks)$}
\psfrag{onramp}[][][.75]{on-ramp $i$}
\psfrag{offramp}[][][.75]{off-ramp $i$}
\psfrag{d}[][][.75]{$d_i(\ks)$}
\psfrag{s}[][][.75]{$s_i(\ks)$}
\psfrag{r}[][][.75]{$e_i(\ks)$}
\includegraphics[width = \linewidth]{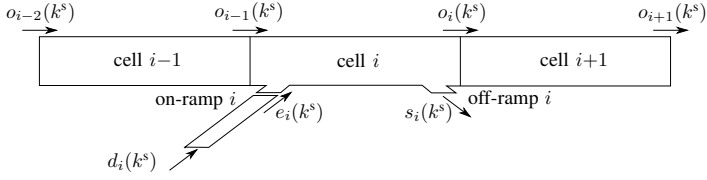}
\caption{The main components and variables used in the ACTM.}
\label{fig:ACTM}
\end{figure}

The model we use in this paper as simulation model and as prediction models   
of the MPC-based controllers is the asymmetric cell transmission model 
(ACTM)\footnote{Note that the approach proposed in this paper is generic, 
and other models could easily be used instead of the ACTM.}  
\cite{Gomes:2006}. 
Figure~\ref{fig:ACTM} illustrates via a stretch of a road with an on-ramp and an off-ramp
the main concepts and variables used by the ACTM. 
The stretch of the road is divided into sections, called \emph{cells}, 
where each cell includes maximally one on-ramp and maximally one off-ramp. 
When a cell contains both an on-ramp and an off-ramp, the on-ramp should be upstream 
of the off-ramp. 
Each cell is identified by an index $i$, where an on-ramp and an   
off-ramp  that belong to a cell adopt the same index as the cell.%

The main variables of the ACTM are:
\begin{itemize}
\item
$\ks$: simulation sampling time step counter 
\item
$n_i(\ks)$: total number of vehicles in cell $i$ for the simulation sampling 
time step $\ks$
\item 
$q_i(\ks)$: total number of vehicles that are queuing on the on-ramp $i$ 
for the simulation sampling time step $\ks$ 
\item
$o_i(\ks)$: mainline outflow of cell $i$/mainline inflow of cell $i+1$ (i.e., 
total number of vehicles that leave cell $i$ towards cell $i+1$)  
for the simulation sampling time step $\ks$
\item 
$s_i(\ks)$: off-ramp outflow of cell $i$, i.e, total number of vehicles that leave cell $i$ via off-ramp $i$ 
for the simulation sampling time step $\ks$
\item
$e_i(\ks)$: on-ramp inflow of cell $i$ (or outflow of on-ramp $i$), i.e., total number of vehicles 
that enter cell $i$ via on-ramp $i$, for the simulation sampling time step $\ks$ 
\item
$\alpha_i(\ks)$: percentage of the vehicles that enter cell $i$ via on-ramp $i$ for the   
simulation sampling time step $\ks$ and can blend with the existing moving flow in the cell, 
i.e., the ACTM assumes two traffic regimes for the vehicles that enter a cell via an on-ramp; 
moving and idling, and correspondingly defines a blending coefficient. 
The vehicles that belong to the idling regime, will stay in the cell during the 
current simulation sampling cycle, while all or part of the vehicles in the moving 
regime may leave the cell during the current simulation sampling cycle. 
The vehicles that are already in the cell at the beginning of 
the current simulation sampling cycle, belong to the moving regime. 
Note that in the limit, when the cell is congested, the moving and the idling regimes will 
merge and adopt the same speed.    
\item
$d_i(\ks)$: demand, given in the number of vehicles per simulation sampling cycle, 
at the beginning of on-ramp $i$ for the simulation sampling time step $\ks$ 
\item 
$\bar{o}_i$: mainline saturation outflow of cell $i$, i.e., maximum number of vehicles 
that can leave cell $i$ towards cell $i+1$ within one simulation sampling cycle 
\item 
$\bar{s}_i$: off-ramp saturation outflow of cell $i$, i.e., maximum number of vehicles 
that can leave cell $i$ via off-ramp $i$  within one simulation sampling  cycle 
\item 
$\bar{n}_i$: capacity of cell $i$, i.e., maximum number of vehicles 
that can be in cell $i$
\item
$\beta_i(\ks)$: percentage of the vehicles that leave cell $i$ for the  simulation time 
step $\ks$ via off-ramp $i$
\item
$\eta^\textrm{m}_i(\ks)$: percentage of the vehicles in the moving regime of cell $i$ 
for the simulation sampling time step $\ks$ that 
can leave this cell within the current simulation sampling cycle, where 
this percentage depends 
on the moving speed the vehicles adopt in the cell for this simulation sampling time step. 
In the limit, when the cell and its neighboring consecutive cells are 
congested, $\eta^\textrm{m}_i(\ks)\hspace*{.15ex}\rightarrow \hspace*{.15ex}0$
\item
$\eta^\textrm{i}_i(\ks)$: percentage of the vacant capacity in cell $i$ 
for the simulation sampling time step $\ks$ that can be 
occupied by the vehicles that enter the cell and join the idling regime 
within one simulation sampling cycle, where this percentage depends on the idling speed 
the vehicles adopt in the cell for this simulation sampling time step 
and the speed of the congestion wave in the cell 
\item
$\xi_i(\ks)$: 
percentage of the vacant capacity in cell $i$ at the beginning of the current 
simulation sampling cycle that can be allocated to the vehicles that 
enter the cell via on-ramp $i$   
\end{itemize}

\begin{figure*}
\psfrag{1}[][][.75]{cell 1}
\psfrag{2}[][][.75]{cell 2}
\psfrag{3}[][][.75]{cell 3}
\psfrag{4}[][][.75]{cell 4}
\psfrag{5}[][][.75]{cell 5}
\psfrag{6}[][][.75]{cell 6}
\psfrag{on2}[][][.75]{on-ramp 2}
\psfrag{off2}[][][.75]{\hspace*{7ex} off-ramp 2}
\psfrag{on4}[][][.75]{on-ramp 4}
\psfrag{off4}[][][.75]{off-ramp 4}
\psfrag{on5}[][][.75]{on-ramp 5}
\psfrag{off5}[][][.75]{\hspace*{7ex} off-ramp 5}
\psfrag{d2}[][][.75]{$d_2$}
\psfrag{d4}[][][.75]{$d_4$}
\psfrag{d5}[][][.75]{$d_5$}
\psfrag{dmain}[][][.75]{\hspace*{-5ex}$d^\textrm{M}$}
\includegraphics[width = \textwidth]{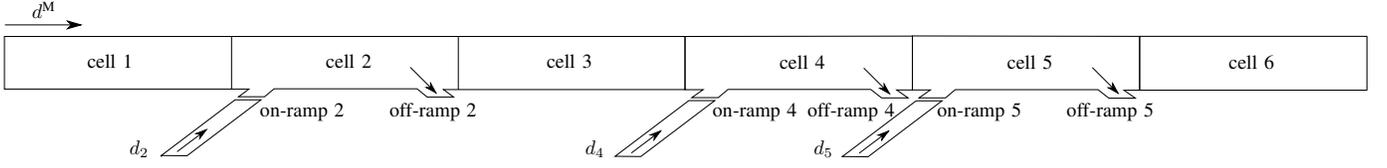}
\caption{Stretch of a single-lane highway used for the case study, which 
has been divided into 6 cells for the ACTM, and includes three metered  
on-ramps and three off-ramps.}
\label{fig:case_study}
\end{figure*}

\begin{remark}
Note that in the given definitions the term ``for the simulation sampling time step $\ks$''  
in the discrete-time domain is an equivalent for the statement 
``during the sampling time interval $\left[\ks \csim , (\ks+1) \csim \right) $'' in 
the continuous-time domain, with $\csim$ the simulation sampling cycle. 
\end{remark}%

\smallskip
The main equations of the ACTM for updating the state variables of 
the model, i.e., $n_i(\ks)$ and $q_i(\ks)$, are given by:
\begin{align}
\label{eq:ACTM_update_n}
n_i(\ks + 1) &= n_i (\ks) + o_{i - 1}(\ks) + e_i(\ks) - o_i(\ks) - s_i(\ks),\\
\label{eq:ACTM_update_6}
q_i(\ks + 1) &= q_i (\ks) + d_{i - 1}(\ks) - e_i(\ks),  
\end{align}
where the mainline outflow of cell $i$ is computed by:
\noindent
\begin{align}
\label{eq:ACTM_mainline_outflow}
o_i(\ks) = \min \bigg\{ &
\left(1-\beta_i(\ks) \right) \left( n_i(\ks) + \alpha_i(\ks) e_i(\ks) \right)\eta_i^\textrm{m}(\ks)  , 
\\
&\left( \bar{n}_{i + 1} -n_{i+1}(\ks) - \alpha_{i+1}  (\ks) e_{i+1}(\ks)  \right)\eta^\textrm{i}_{i+1}(\ks) , 
\nonumber\\
&\bar{o}_i, 
\frac{\dps 1-\beta_i(\ks)}{\beta_i(\ks)} \bar{s}_i  
\bigg\}.  \nonumber
\end{align} 
Note that in the ACTM, the cells are defined in such a way that 
the mainline inflow, $o_{i-1}(\ks)$, does not contribute to the mainline outflow, $o_i(\ks)$,  
of cell $i$ (see \eqref{eq:ACTM_mainline_outflow}), while a percentage of the on-ramp inflow, 
i.e., $\alpha_i(\ks) e_i(\ks)$, contributes to the mainline outflow of the cell. 
In other words, all the vehicles in the mainline inflow of cell $i$ 
blend with the idling flow in the cell.   
The on-ramp inflow of cell $i$ is computed by: 
\begin{align}
\label{eq:e_i}
e_i(\ks) = 
\left\{
\begin{array}{l}
\min\bigg\{
q_i(\ks) + d_i(\ks) , \xi_i \left( \bar{n}_i  - n_i(\ks) \right)
\bigg\},   
\\
\textrm{\hspace*{12ex} if the on-ramp is not metered} 
\\
\\
\min\bigg\{
q_i(\ks) + d_i(\ks) , \xi_i \left( \bar{n}_i  - n_i(\ks) , \mu_i(\ks) \right) 
\bigg\},  
\\
\textrm{\hspace*{12 ex} if the on-ramp is metered}
\end{array},
\right.
\end{align}
where $\mu_i(\ks)$ is the metering rate of the metered on-ramp $i$ for the simulation sampling 
time step $\ks$, i.e., total number of vehicles that are allowed to enter 
cell $i$ via on-ramp $i$  at the  simulation sampling time step $\ks$.  

\noindent
Finally, for the off-ramp outflow of cell $i$, we have\footnote{In the boundary case 
where $\beta_i(\ks) = 1$, the following equation may be used: 
$
s_i(\ks) = \min \bigg\{ \bar{s}_i , \left( n_i(\ks) + 
\alpha_i(\ks) e_i(\ks) \right) \eta^\textrm{m}_i(\ks) \bigg\}.
$}:
\begin{align}
s_i(\ks) = \frac{\dps \beta_i(\ks)}{1 - \beta_i(\ks)} o_i(\ks).
\end{align}

The stretch of a single-lane highway we use for the case study is illustrated 
in Figure~\ref{fig:case_study}. 
The highway stretch is divided into 6 cells where the second, 
fourth, and fifth cells each include one metered on-ramp and one off-ramp. 
The traffic flow is in the direction of the increase in the cell indices, 
i.e, it is from left to right in Figure~\ref{fig:case_study}. 
The simulation parameters have been selected based on   
\cite{Munoz:2004} and in line with \cite{Gomes:2006}. 
These parameters are listed in Table~\ref{table:simulation_parameters}. 
We run the simulations for 180 simulation time steps, i.e., for a period of 1 hours.%

\begin{figure}
\centering
\includegraphics[width = .75\linewidth]{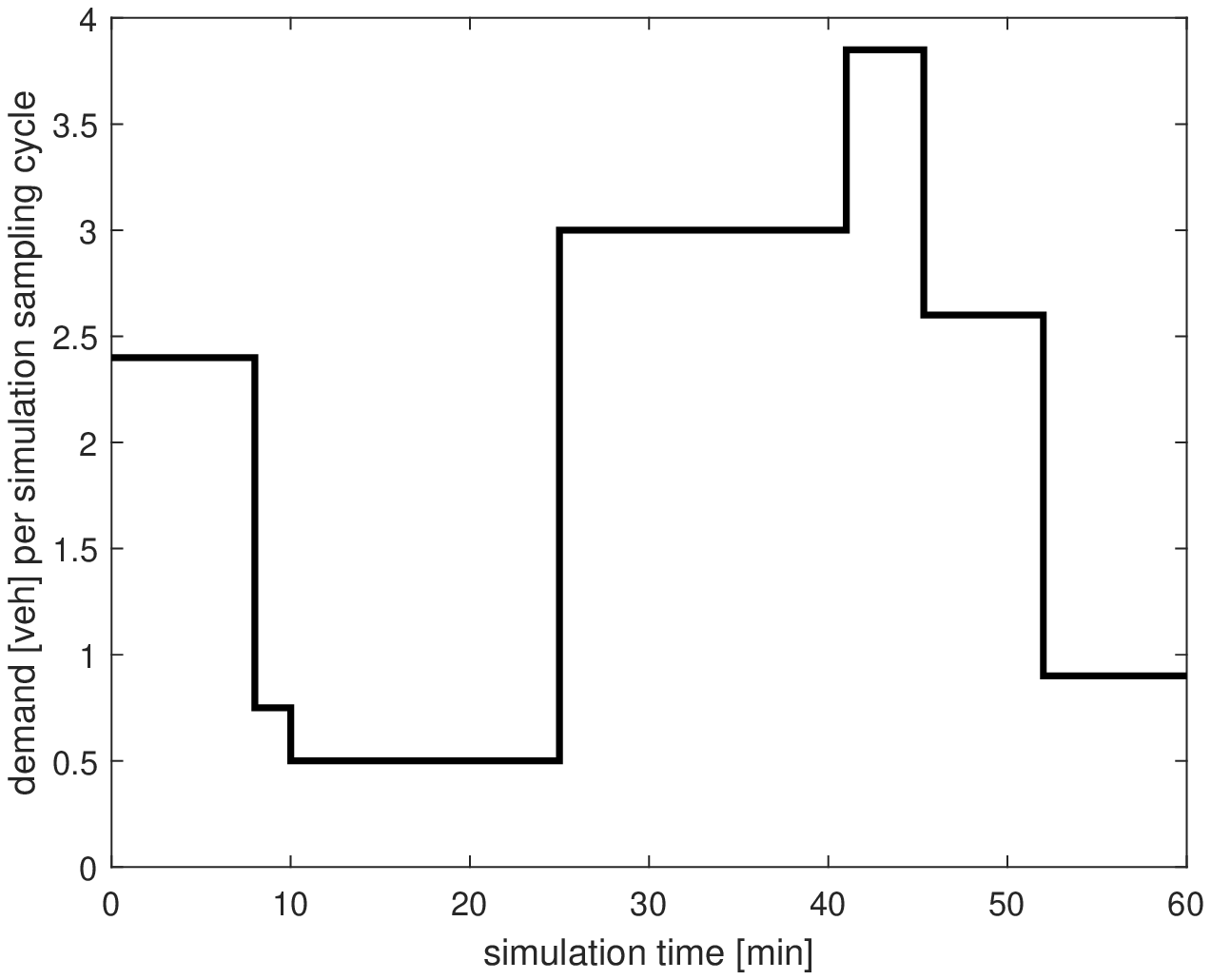}
\vspace*{3ex}
    
\includegraphics[width = .75\linewidth]{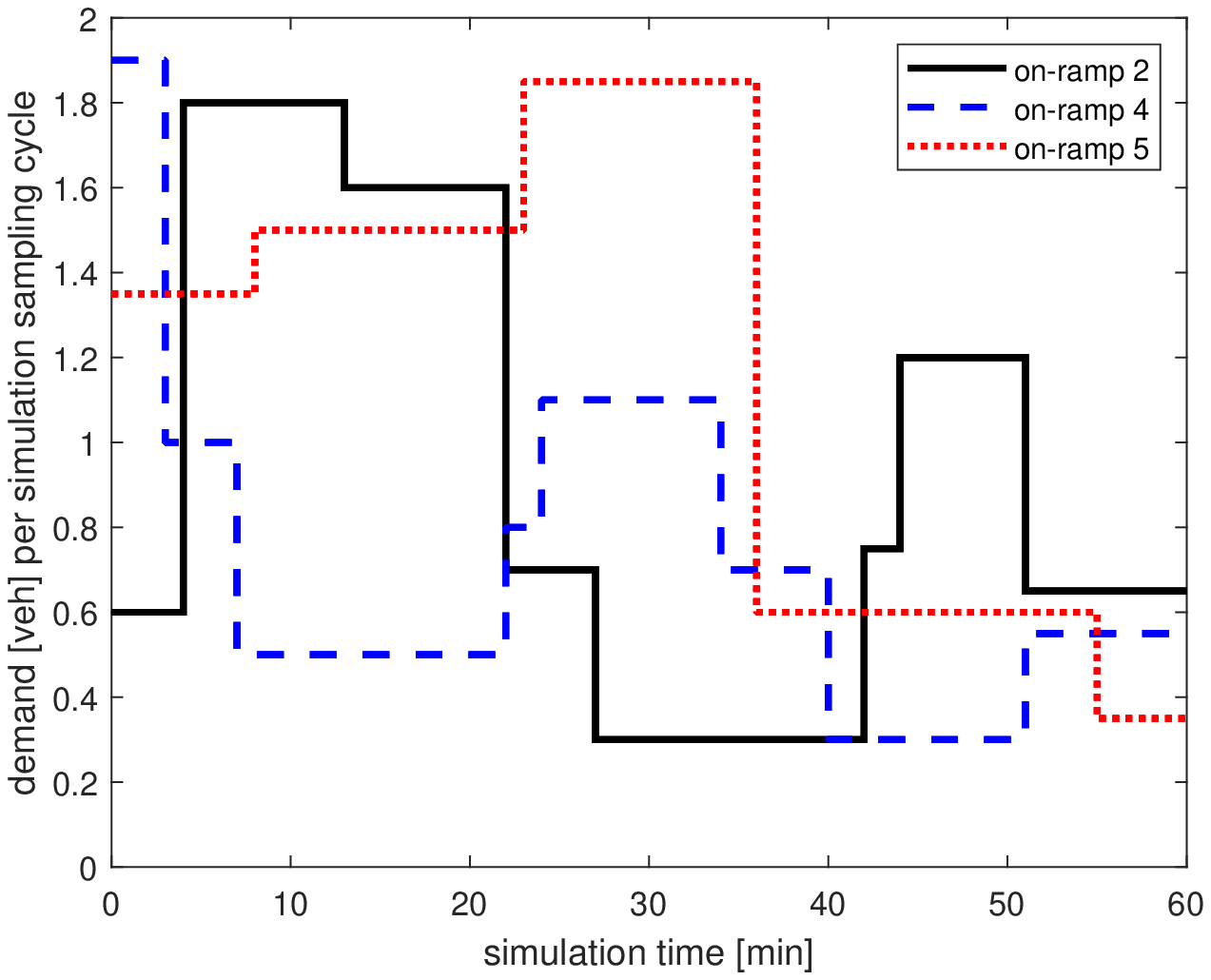}
\caption{Demand profiles of the mainstream road (top plot) 
and of the on-ramps (bottom plot) used for the case study. 
\label{fig:demands}}
\end{figure}

The measured values of the demands at the mainstream and on-ramps 
used for the case study are represented in Figure~\ref{fig:demands}. 
In order to make the scenarios of the case study more realistic, 
for the predicted values of the demands at the origin of the 
mainstream, i.e., $d^\textrm{M}$, and at the beginning of the on-ramps, 
i.e., $d_2$, $d_4$, $d_5$, we randomly add/subtract a random error of 
up to 10\% of the measured values to/from them. 
The initial states of the network (given in [veh]) are 
$n_1(0) = 32.6,\ n_2(0) = 36.2,\ n_3(0)= 5.1,\ n_4(0) = 25.3,\ n_5(0)= 3.9,\ 
q_2(0) = 5.5,\ q_4(0) = 9.6,\ q_5(0) = 1.6$.

\begin{savenotes}
\begin{table*}
\setlength\tabcolsep{3pt} 
\centering
\caption{Simulated values of the parameters used for the ACTM and for the control 
   system in the case study.
   \label{table:simulation_parameters}} 
\begin{tabular}{|l|c|c|c|c|c|c|}
\hline
\cellcolor{gray!25}parameter name  & simulation and control sampling cycle & 
cell length & free-flow speed & 
idling speed  & average vehicle length\footnote{The given value for the average vehicle length 
 is in line with the findings of a project performed in 2015 by 
the New England Section of the Institute of Transportation Engineers Technical 
Committee, which is available via \url{http://neite.org/Documents/Technical/}.} 
& cell capacity \\ 
\hline
\cellcolor{gray!25}simulated value  & $20$~[s] & $560$~[m] & $28$~[m/s] & $6.5$~[m/s] 
& $7$~[m] & $80$~[veh]\\ 
\hline
\end{tabular}

\vspace*{2ex}

\begin{tabular}{|l|c|c|c|c|c|c|c|c|c|c|c|c|c|c|}
\hline
\cellcolor{gray!25}parameter notation  &  $\alpha_2$ &  
$\alpha_4$ &  $\alpha_5$ & $\bar{o}$ & 
$\bar{s}$ & $\beta_2$ & $\beta_4$ & $\beta_5$ 
& $\eta^\textrm{m}_2$ & $\eta^\textrm{m}_4$ & $\eta^\textrm{m}_5$ &
$\eta^\textrm{i}$& $\xi$ & $\rho^\textrm{crit}$ \\
\hline
\cellcolor{gray!25}simulated value & 0.6~[-] & 0.8~[-] & 0.7~[-] &  
$8$~[veh] & $6$~[veh] & 0.35~[-] & 0.62~[-] & 0.43~[-] & 
0.8~[-] & 0.65~[-] & 0.8~[-] & 0.3~[-] &  0.4~[-]  & $0.0335$~[veh/m/lane] \\
\hline
\end{tabular}

\end{table*}
\end{savenotes} 

\subsection{Base and parallel controllers}

 \begin{figure}
 \centering
 \psfrag{ev}[][][0.8]{Evaluate}
 \psfrag{10}[][][0.8]{\eqref{eq:Alinea}}
 \psfrag{ANN}[][][0.8]{ANN}
 \psfrag{Alinea}[][][0.8]{ALINEA}
 \psfrag{high}[][][0.8]{Highway (ACTM with disturbed external inputs)}
 \psfrag{ACTM}[][][0.8]{ACTM}
 \psfrag{pmpc1}[][][0.8]{PMPC(1)}
 \psfrag{pmpc2}[][][0.8]{PMPC(2)}
 \psfrag{cmpc1}[][][0.8]{CMPC(1)}
 \psfrag{cmpc2}[][][0.8]{CMPC(2)}
 \psfrag{base}[][][0.8]{Base block}
 \psfrag{par}[][][0.8]{Parallel block}
 \psfrag{theta}[][][0.8]{$\bm{\theta}^\textrm{ALINEA}(k)$}
 \psfrag{mu}[][][0.8]{$\bm{\mu}(k)$}
 \psfrag{state }[][][0.8]{states and external inputs}
 \psfrag{mutildei}[][][0.8]{$\tilde{\bm{\theta}}^\textrm{BCi}(k;\np(1)) , \tilde{\bm{\theta}}^\textrm{BCi}(k;\np(2))$}
 \psfrag{mutildee}[][][0.8]{$\tilde{\bm{\mu}}^\textrm{BCe}(k ;\np(1)) , \tilde{\bm{\mu}}^\textrm{BCe}(k; \np(2))$}
 \psfrag{mup1}[][][0.8]{$\tilde{\bm{\mu}}^\textrm{PMPC(1)}(k;\bar{n}^\textrm{e})$}
 \psfrag{mup2}[][][0.8]{$\tilde{\bm{\mu}}^\textrm{PMPC(2)}(k;\bar{n}^\textrm{e})$}
 \psfrag{muc2}[][][0.8]{$\tilde{\bm{\mu}}^\textrm{CMPC(2)}(k;\bar{n}^\textrm{e})$}
 \psfrag{muc1}[][][0.8]{$\tilde{\bm{\mu}}^\textrm{CMPC(1)}(k;\bar{n}^\textrm{e})$} 
 \psfrag{muANN}[][][0.8]{$\tilde{\bm{\mu}}^\textrm{ANN}(k;\bar{n}^\textrm{e})$}
 \psfrag{muA}[][][0.8]{$\tilde{\bm{\mu}}^\textrm{ALINEA}(k;\bar{n}^\textrm{e})$}
 \psfrag{select}[][][0.8]{Selector}
 \psfrag{tttp1}[][][0.7]{$J^\textrm{PMPC(1)}(k)$}
 \psfrag{tttp2}[][][0.7]{$J^\textrm{PMPC(2)}(k)$}
 \psfrag{tttc2}[][][0.7]{$J^\textrm{CMPC(2)}(k)$}
 \psfrag{tttc1}[][][0.7]{$J^\textrm{CMPC(1)}(k)$} 
 \psfrag{tttANN}[][][0.7]{$J^\textrm{ANN}(k)$}
 \psfrag{tttA}[][][0.7]{$J^\textrm{ALINEA}(k)$}
 \psfrag{mubest}[][][0.8]{$\bm{\mu}^\textrm{best}(k)$}
\includegraphics[width = \linewidth]{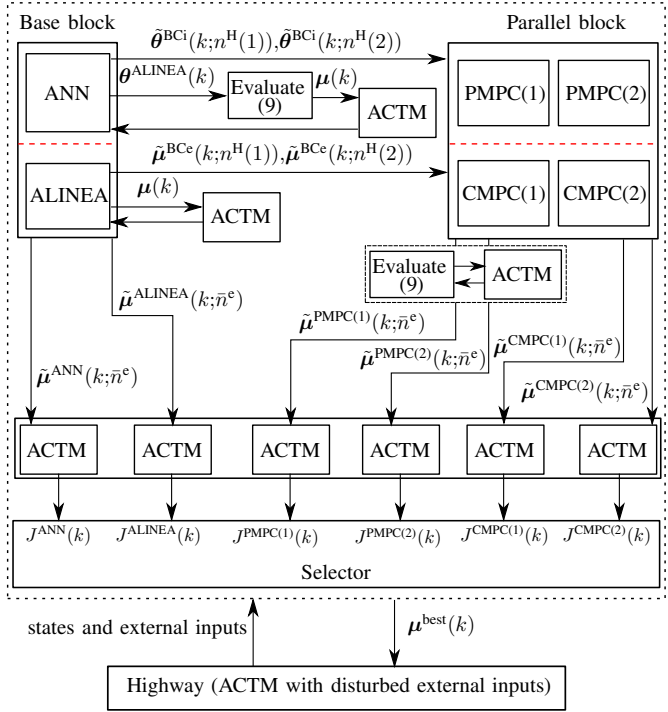}
 \caption{ The base-parallel integrated control architecture 
 implemented to the stretch of the highway for the case study.} 
 \label{fig:case_study_controller}
 \end{figure}

The base-parallel control architecture we have used for this case study 
is illustrated in Figure~\ref{fig:case_study_controller}.  
In the base block, we have considered an explicit (ALINEA) and an implicit (ANN)  
base controller, which are detailed next. 
ALINEA \cite{Alinea:1991} is a feedback-based control approach for ramp metering  
in road traffic. 
The control signal produced by ALINEA for the control sampling time step $k$ for  
the ramp meter of on-ramp $i$ is
\begin{equation}
\label{eq:Alinea}
\mu_i(k) = \max\Big\{  
\mu_i(k - 1) + \theta^\textrm{ALINEA}_i(k) \left( \rho^\textrm{crit}_i - \rho_i(k) \right)
 , 0\Big\}, 
\end{equation}
with $\theta^\textrm{ALINEA}_i(k)$ the tuning parameter of ALINEA for cell $i$ 
at the control sampling time step $k$, and 
$\rho^\textrm{crit}_i$ and $\rho_i(k)$ the critical and measured densities 
(i.e., total number of vehicles per unit length per lane) 
of cell $i$ at the control sampling time step $k$ directly downstream of on-ramp $i$. 
For the ALINEA block, we apply the constant gain of $\theta^\textrm{ALINEA} = 0.016$  
from \cite{Papageorgiou:1990}, when using the SI units for the variables in \eqref{eq:Alinea}. 
Moreover, the values (given in [veh] per simulation sampling cycle) 
$\mu_2= 0.5, \ \mu_4 = 0.2,$ and $\mu_5 = 0.4$ are used as the previous control signals 
in order to evaluate the control signal of ALINEA for the initial simulation sampling time step.  
For  $\rho^\textrm{crit}_i$, we use $0.0335$ veh/m/lane \cite{Hegyi:2005} 
for all the 6 cells shown in Figure~\ref{fig:case_study}. 
Note that since $\rho_i(k)$ is not computed directly as a state variable via the ACTM, 
we should compute it at every control sampling time step in veh/m/lane, which based on the  
definition of the density, is determined by $\rho_i(k)=n_i(k)/(\textrm{cell length})$, 
reminding that the highway of the case study is single-lane. 
In Figure~\ref{fig:case_study_controller}, the vector $\bm{\mu}(k)$ includes 
$\mu_2(k)$, $\mu_4(k)$, and $\mu_5(k)$, i.e., the ALINEA block  
illustrated in Figure~\ref{fig:case_study_controller} includes the 
 control policy of \eqref{eq:Alinea} and hence, the tuned parameters of ALINEA 
for all the three metered on-ramps.%

\begin{figure}
\centering
\psfrag{ANN}[][][.8]{ANN for cell $i$}
\psfrag{q}[][][.8]{$q_i(k)$}
\psfrag{n}[][][.8]{$n_i(k)$}
\psfrag{theta}[][][.8]{$\theta^\textrm{ALINEA}_i(k)$}
\psfrag{d}[][][.8]{$d_i(k)$}
\psfrag{o}[][][.8]{$o_{i-1}(k)$}
\includegraphics[width = .55\linewidth]{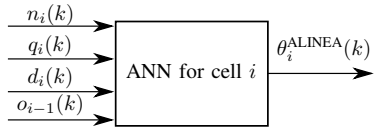}

\vspace*{2ex}
\includegraphics[width =  \linewidth]{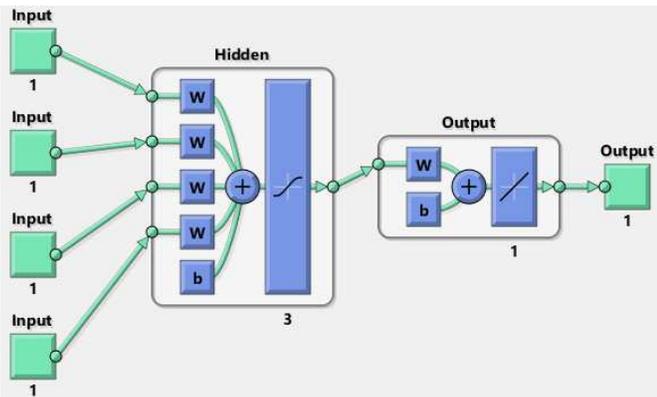}
\caption{The ANN-based mapping used as an implicit base controller per cell for the case study.}
\label{fig:implicit_ANN}
\end{figure}

For the implicit base controller, a mapping based on an artificial neural network 
(ANN) is trained (see Figure~\ref{fig:implicit_ANN}), which receives the state variables 
$n_i(k)$ and $q_i(k)$ and the  uncontrolled external inputs $d_i(k)$ and $o_{i-1}(k)$ 
of a cell for every control sampling time step $k$, 
and produces a value for $\theta^\textrm{ALINEA}_i(k)$ in \eqref{eq:Alinea}. 
The ANN is a feedforward network with one hidden layer of size three. 
To train this mapping, we have generated a dataset of size $500$, including 
$n_i$, $q_i$, $d_i$, and $o_{i-1}$ with $i = 2,4,5$ for $500$ various traffic 
scenarios as the inputs of the ANN for cell $i$, and optimized values of 
$\theta^\textrm{ALINEA}_i(k)$ as the output. 
The input values are considered such that a wide range of possible scenarios 
is covered. To generate the output values, we consider 
each metered cell as an isolated one, i.e., with infinite capacity for 
the leaving flows, and minimize the difference between the critical density 
$\rho^\textrm{crit}_i$ and the cell's expected density $\rho^\textrm{exp}_i(k+1)$ 
with respect to $\theta^\textrm{ALINEA}_i(k)$ as the optimization variable. 
This is because the objective of ALINEA 
is to keep the road's density downstream of an on-ramp near the critical density 
(see \eqref{eq:Alinea} and \cite{vandenWeg:2019}). 
The expected density is determined by dividing the expected total number 
of vehicles $n^\textrm{exp}_i(k+1)$ in cell $i$ by the length of the cell, with
\begin{align*}
n^\textrm{exp}_i(k+1) = n_i(k) + o_{i - 1}(k) + e_i(k) - o_i(k) - s_i(k),
\end{align*} 
where $e_i(k)$ is computed by the second formulation of \eqref{eq:e_i} and $\mu_i(k)$ 
by \eqref{eq:Alinea}. 
The generated dataset is divided into a training dataset including 400 data, 
and a validation dataset including the rest 100 data. 
Note that the ANN in Figure~\ref{fig:case_study_controller} includes three mappings for the 
three ramp meters, i.e., the vector $\bm{\theta}^\textrm{ALINEA}(k)$ includes 
$\theta^\textrm{ALINEA}_2(k)$, $\theta^\textrm{ALINEA}_4(k)$, and $\theta^\textrm{ALINEA}_5(k)$. 
For the case study, in addition to the initial state of the network and 
the initial values of the measured demands, we have used 
$o_1(0) = 3.8,\ o_3(0) = 3.2,$ and $o_4(0) = 0.6$ (given in [veh] per simulation sampling cycle) 
to evaluate the outputs of the three ANN-based mappings for the initial simulation 
sampling time step. 
The output $\bm{\theta}^\textrm{ALINEA}(k)$ of the ANN is first transformed into 
$\bm{\mu}(k)$ using \eqref{eq:Alinea} (see Figure~\ref{fig:case_study_controller}). 
The ACTM is then used to generate a sequence $\tilde{\bm{\mu}}$ 
using  the control inputs 
generated by the ALINEA and the ANN blocks.%

For each base controller, we consider a parallel cell with two MPC-based 
controllers, where for the explicit base controller (ALINEA), the MPC-based 
controllers (CMPC(1) and CMPC(2) in Figure~\ref{fig:case_study_controller}) 
are formulated using conventional MPC, and for the implicit base controller (ANN), 
parameterized MPC (PMPC(1) and PMCP(2) in Figure~\ref{fig:case_study_controller}) 
is used. 
We use the ``fmincon'' solver with multiple starting optimization points and 
the Sequential Quadratic Programming (SQP) algorithm from the Matlab optimization  
toolbox, to solve the MPC optimization problems, 
which, in general, are nonlinear and non-convex. 
The function tolerance and the step tolerance are set to, respectively, $10^{-3}$ and $10^{-7}$. 
The cost function $J(k)$ of the MPC optimization problems at the control sampling 
time step $k$ is defined by
\begin{equation}
\label{eq:MPC_objective_function}
J(k) = \textrm{TT}(k) - \gamma \textrm{TD}(k),
\end{equation} 
with $\textrm{TT}$ and $\textrm{TD}$ the travel time and the traveled distance 
of all the vehicles. We have used $\gamma = 0.8$ for the simulations.  
The MPC optimization problems are subject to constraints, including the traffic  
dynamics, upper and lower bounds for the ramp meters, and lower bound for the 
queue lengths on the on-ramps.%

The main difference between the two MPC-based controllers 
indexed by 1 and 2 in each cell 
is in the size of their prediction horizons $\np(1)$ and $\np(2)$, 
i.e., 3 and 10. 
More specifically, CMPC(1) and PMPC(1) have the prediction horizon $\np(1)$ 
and CMPC(2) and PMPC(2) have the prediction horizon $\np(2)$. 
Note that a large prediction horizon 
provides a more extensive vision of the future, which may help to 
reduce the negative effects of the finite horizon of MPC on the degree of optimality 
of the solutions. 
However, the cumulative errors resulting from the inaccuracies in the prediction model 
are larger for a larger prediction horizon.%

The candidate control inputs produced by ANN, ALINEA, PMPC(1), PMPC(2), CMPC(1), and CMPC(2) 
are all sent to the evaluation block (see Figure~\ref{fig:case_study_controller}), 
where six ACTM blocks will in parallel estimate the realized value 
of the cost function of all the  vehicles for a prediction time horizon of size 
$\bar{n}^\textrm{e}$ (assuming that  
$\bar{n}^\textrm{e}\leq \np(1) < \np(2)$) for each candidate control input. 
We have used $\bar{n}^\textrm{e} = 3$ for the case study. 
The ``Selector'' compares these values and selects the control input $\bm{\mu}^\textrm{best}(k)$ 
that corresponds to the minimum cost.%

\subsection{Results}

\begin{table*}
\centering
\caption{The results of a 1-hour simulation for the case study.\label{table:results}} 
\begin{tabular}{|c|c|c|c|}
\hline
\cellcolor{gray!25}control approach  & \cellcolor{gray!25}$J^\textrm{total} = 
{\dps\sum_k}\ \left(\textrm{TT}(k) - \gamma \textrm{TD}(k) \right)$~[h] & 
\cellcolor{gray!25} $n^\textrm{total} $~[veh] 
& \cellcolor{gray!25}average cost per vehicle~[s] \\ 
\hline
Alinea & 455.95 & 6.1342 $\times 10^4$ & 26.76 \\
\hline
ANN    & 431.71 & 6.4531 $\times 10^4$ & 24.08 \\
\hline
CMPC(1) & 412.76 & 6.7738 $\times 10^4$ & 21.94 \\
\hline
CMPC(2) & 412.76 & 6.7738 $\times 10^4$ & 21.94 \\
\hline
PMPC(1) & 412.91 & 6.7710 $\times 10^4$ & 21.95 \\
\hline
PMPC(2) & 412.91 & 6.7679 $\times 10^4$ & 21.96 \\
\hline
{\color{blue}Base-parallel architecture} & {\color{blue}412.69} & {\color{blue}6.7582 $\times 10^4$} 
& {\color{blue}21.98} \\
\hline
\end{tabular}
\end{table*}

Table~\ref{table:results} includes the results of the case study for a 
1-hour simulation repeated with different control approaches,  
Alinea, ANN, CMPC(1), CMPC(2), PMPC(1), PMPC(2), and the base-parallel 
control architecture given by Figure~\ref{fig:case_study_controller}. 
Since, in general, the optimization problems of the MPC-based controllers 
are non-convex, the optimization solvers may return a locally optimal solution 
instead of a global optimum. In order to deal with this problem, we have 
proposed three different starting points for the MPC optimization solvers, i.e., 
\begin{itemize}
\item
solution of the optimization corresponding to the previous control sampling 
time step, ignoring the first element of the sequence 
and repeating the last element twice;  
\item
average of the solution of the optimization corresponding to the 
previous control sampling time step (ignoring the first element of 
the sequence and repeating the last element twice) and the solution 
of the optimization corresponding to the second previous control 
sampling time step (ignoring the first two elements of the sequence and 
repeating the last element three times), 
where this starting point can be used for $k\geq 2$;
\item
average of the solution of the optimization corresponding to all the 
previous control sampling time steps (ignoring those elements of 
the sequences that correspond to the past control sampling time 
steps and repeating the last elements to cover the prediction time 
window).
\end{itemize}%
In Table~\ref{table:results} for each control approach, we have represented 
the realized value of the total cost function, $J^\textrm{total}$, in [h] for 
the 1-hour simulation period, which has been computed via 
\eqref{eq:MPC_objective_function} per control 
sampling time step and accumulated across the simulation time window.  
The total number of vehicles, $n^\textrm{total}$, 
that can travel through the controlled stretch of the road within the 
fixed simulation time window is another indication for the control performance. 
In other words, we expect a high-performing control approach to minimize 
$J^\textrm{total}$, while allowing higher numbers of vehicles that intend to 
enter the stretch of the road to travel via it in the given simulation  time window. 
Therefore, the ratio of these two values, i.e., the average cost per vehicle, 
can be an indication of the control performance. The lower this ratio, the better 
the control approach. 
Hence, in Table~\ref{table:results} we have also represented $n^\textrm{total}$ 
in [veh] and the average cost per vehicle in [s] (i.e., $J^\textrm{total}*3600$ divided by 
$n^\textrm{total}$), for all the control approaches used in the case study.%

Based on the results given in Table~\ref{table:results}, CMPC(1), CMPC(2), 
PMPC(1), and PMPC(2) result in the lowest values for the  average cost per vehicle.  
Comparing the three real-time control approaches (i.e., they can always produce their 
candidate control inputs in a time budget smaller than or equal to the control sampling 
time of the controlled system) ALINEA, ANN, and the proposed base-parallel control 
architecture, the lowest average cost per vehicle corresponds to the proposed base-parallel 
control architecture. 
It is also important to indicate that among all the 7 control approaches 
considered in this case study, the proposed base-parallel control 
approach results in the lowest value of the total cost.%


\section{Discussion}
\label{sec:discussion}

The conventional and parameterized MPC approaches have resulted in the best 
performance among the other control methods used 
in the case study. This is somehow expected since 
MPC approaches minimize the cost function at every control 
sampling time step by solving an online optimization problem, 
based on the updated measured values of the state and uncontrolled 
external inputs. 
The main issue with MPC, however, is the high computation time, 
which may exceed the control sampling cycle of the controlled system, especially when the scales 
of the controlled system and the complexity of the dynamics increase. 
Moreover, due to the non-convexity of the optimization problem, several starting 
optimization points should be considered to make sure more sub-regions in the 
optimization search region are covered. This will cause increased computational 
burden and computation time.%

However, when the MPC-based approaches are used in the proposed parallel block, 
the issue with the computational burden and computation time can be tackled  
due to the following reasons:
\begin{compactitem}
\item
The MPC-based controllers are given a time budget that is smaller than or equal to 
the control sampling time of the controlled system. 
Therefore, the control system will not wait longer than this time budget, and 
in case the control inputs returned by any of the MPC-based controllers result 
in a non-satisfactory performance, they will be excluded by the evaluation block 
and the selector. 
\item
Several MPC-based controllers will run in parallel, i.e., instead of running one 
MPC module in a serial set-up with various optimization starting points (as for 
CMPC(1), CMPC(2), PMPC(1), and PMPC(2)), they will be run in parallel. 
Therefore, the computational burden is distributed among the parallel MPC-based 
controllers, which results in reduced computation time. 
\item
The existence of the base block next to the online MPC-based controllers provides 
a warm-start for them that reduces the computational burden and the computation 
time, which are large for the conventional and parameterized MPC-based controllers, 
CMPC(1), CMPC(2), PMPC(1), and PMPC(2), due to an extensive 
exploration of the optimization search region. 
\end{compactitem}
Moreover, a detailed assessment of the prediction horizon for real-life processes is 
not possible prior to running the control procedure, particularly 
when the nonlinearities increase. Therefore, we allow both a small and a large prediction 
horizon for the MPC-based controllers via the use of parallel cells within the 
proposed control architecture.%

The lower value of the realized total cost function corresponding 
to the base-parallel control architecture in comparison with the MPC-based approaches 
that aim at minimizing the same cost function online at every control 
sampling time step, together with the fact that for the majority of the control 
sampling time steps, one of the MPC-based controllers has won the competition 
in the evaluation block, may be explained via the following reason. 
The warm-start provided via the base block for the online MPC optimization problems 
in the parallel block, has improved the performance of these controllers for several control 
sampling time steps. 
In other words, although we may provide several starting points for the online MPC optimization 
problems, this will not guarantee that the entire optimization search region 
will necessarily be covered. 
More specifically, this will not guarantee that the sub-regions that correspond to 
global or better local optima will be covered, while the warm-start provided by the 
base block has in several cases allowed the online optimizers to search such sub-regions.%

\section{Conclusions and Future Work}
\label{sec:conclusions}

We have proposed an integrated base-parallel control architecture 
with the aim of addressing the current gap in efficient real-time 
implementation of MPC-based control approaches for highly nonlinear 
systems with fast dynamics and a large number of control constraints. 
In this architecture, several offline tuned or optimized controllers 
and online optimization-based controllers can run in parallel.%

We have performed a simulation for controlling the metered on-ramps of 
a stretch of a highway. The simulation results have shown that, among the 
control approaches that can perform in real time, the proposed base-parallel  
control architecture has the performance (considering the average cost 
per vehicle) that is the closest to the online MPC-based 
controllers, which produce optimal control inputs, but suffer from 
high computation times. 
Moreover, the least value of the overall cost among all these 7 control 
approaches, corresponds to the base-parallel control architecture. 
The new control architecture results in a control system that is very 
flexible and its architecture can easily be changed or modified online.%

Relevant topics that require further exploration in the future include:
\begin{itemize}
\item
Stability analysis of the proposed base-parallel control architecture, 
assuming individually stable base and parallel controllers. 
\item
More extensive simulations including various large-scale controlled 
systems with highly nonlinear dynamics, and including different base 
and parallel controllers. 
\item
In-depth study and analysis of the influences of different tuning parameters of MPC 
for various dynamics of the controlled system. This can be done by running 
several parallel MPC-based controllers in the parallel block and tracking the 
trend of the selection of each candidate control input based on the 
ongoing dynamics of the controlled system.%
\end{itemize}


\section*{Acknowledgments}

This research has been supported by the NWO-NSFC project ``Multi-level 
predictive traffic control for large-scale urban networks'' 
(629.001.011), which is partly financed by the Netherlands 
Organization for Scientific Research (NWO).%




\begin{thebibliography}{10}
\providecommand{\url}[1]{#1}
\csname url@samestyle\endcsname
\providecommand{\newblock}{\relax}
\providecommand{\bibinfo}[2]{#2}
\providecommand{\BIBentrySTDinterwordspacing}{\spaceskip=0pt\relax}
\providecommand{\BIBentryALTinterwordstretchfactor}{4}
\providecommand{\BIBentryALTinterwordspacing}{\spaceskip=\fontdimen2\font plus
\BIBentryALTinterwordstretchfactor\fontdimen3\font minus
  \fontdimen4\font\relax}
\providecommand{\BIBforeignlanguage}[2]{{%
\expandafter\ifx\csname l@#1\endcsname\relax
\typeout{** WARNING: IEEEtran.bst: No hyphenation pattern has been}%
\typeout{** loaded for the language `#1'. Using the pattern for}%
\typeout{** the default language instead.}%
\else
\language=\csname l@#1\endcsname
\fi
#2}}
\providecommand{\BIBdecl}{\relax}
\BIBdecl

\bibitem{Maciejowski:2002}
J.~Maciejowski, \emph{Predictive Control with Constraints}.\hskip 1em plus
  0.5em minus 0.4em\relax London, UK: Prentice Hall, 2002.

\bibitem{Huyck:2012}
B.~Huyck, H.~J. Ferreau, M.~Diehl, J.~{De Brabanter}, J.~F.~M. {Van Impe},
  B.~{De Moor}, and F.~Logist, ``Towards online model predictive control on a
  programmable logic controller: {P}ractical considerations,''
  \emph{Mathematical Problems in Engineering}, vol. 2012, pp. 1--20, 2012.

\bibitem{Huyck:2014}
B.~Huyck, J.~{De Brabanter}, J.~F. {Van Impe}, and F.~Logist, ``Online model
  predictive control of industrial processes using low level control hardware:
  {A} pilot-scale distillation column case study,'' \emph{Control Engineering
  Practice}, vol.~28, pp. 34--48, 2014.

\bibitem{Wang:2010}
Y.~Wang and S.~Boyd, ``Fast model predictive control using online
  optimization,'' \emph{IEEE Transactions on Control Systems Technology},
  vol.~18, no.~2, pp. 267--278, 2010.

\bibitem{Houska:2011}
B.~Houska, H.~J. Ferreau, and M.~Diehl, ``An auto-generated real-time iteration
  algorithm for nonlinear {MPC} in the microsecond range,'' \emph{Automatica},
  vol.~47, no.~10, pp. 2279--2285, 2011.

\bibitem{Bemporad:2002}
A.~Bemporad, M.~Morari, V.~Dua, and E.~N. Pistikopoulos, ``The explicit linear
  quadratic regulator for constrained systems,'' \emph{Automatica}, vol.~38,
  no.~1, pp. 3--20, 2002.

\bibitem{Johansen:2002}
T.~A. Johansen, ``On multi-parametric nonlinear programming and explicit
  nonlinear model predictive control,'' in \emph{41st IEEE Conference on
  Decision and Control}, Las Vegas, USA, December 2002, pp. 2768--2773.

\bibitem{Bemporad:2006}
A.~Bemporad and C.~Filippi, ``An algorithm for approximate multiparametric
  convex programming,'' \emph{Computational Optimization and Applications},
  vol.~35, no.~1, pp. 87--108, 2006.

\bibitem{Bemporad:2009}
A.~Alessio and A.~Bemporad, \emph{Nonlinear Model Predictive Control Lecture
  Notes in Control and Information Sciences}.\hskip 1em plus 0.5em minus
  0.4em\relax Berlin, Germany: Springer, 2009, vol. 384, ch. A Survey on
  Explicit Model Predictive Control.

\bibitem{Zeilinger:2011}
M.~N. Zeilinger, C.~N. Jones, and M.~Morari, ``Real-time suboptimal model
  predictive control using a combination of explicit {MPC} and online
  optimization,'' \emph{IEEE Transactions on Automatic Control}, vol.~56,
  no.~7, pp. 1524--1534, 2011.

\bibitem{Zegeye:2012}
S.~K. Zegeye, B.~{D}e Schutter, J.~Hellendoorn, E.~A. Breunesse, and A.~Hegyi,
  ``A predictive traffic controller for sustainable mobility using
  parameterized control policies,'' \emph{IEEE Transactions on Intelligent
  Transportation Systems}, vol.~13, no.~3, pp. 1420--1429, 2012.

\bibitem{Muehlebach:2016}
M.~Muehlebach and R.~D'Andrea, ``Parameterized infinite-horizon model
  predictive control for linear time-invariant systems with input and state
  constraints,'' in \emph{American Control Conference (ACC)}, Boston, USA, July
  2016, pp. 2669--2674.

\bibitem{Cagienard:2007}
R.~Cagienard, P.~Grieder, E.~C. Kerrigan, and M.~Morari, ``Move blocking
  strategies in receding horizon control,'' \emph{Journal of Process Control},
  vol.~17, no.~6, pp. 563--570, 2007.

\bibitem{Shekhar:2015}
R.~C. Shekhar and C.~Manzie, ``Optimal move blocking strategies for model
  predictive control,'' \emph{Automatica}, vol.~61, pp. 27--34, 2015.

\bibitem{Grune:2011}
L.~Gr\"{u}ne and J.~Pannek, \emph{Nonlinear Model Predictive Control: Theory
  and Algorithms}.\hskip 1em plus 0.5em minus 0.4em\relax London, UK:
  Springer-Verlag, 2011.

\bibitem{Bemporad:2003}
A.~Bemporad and C.~Filippi, ``Suboptimal explicit receding horizon control via
  approximate multiparametric quadratic programming,'' \emph{Journal of
  Optimization Theory and Applications}, vol. 117, no.~1, pp. 9--38, 2003.

\bibitem{Chen:2000}
W.~H. Chen, D.~J. Ballance, and J.~O'Reilly, ``Model predictive control of
  nonlinear systems: Computational burden and stability,'' \emph{IEE
  Proceedings - Control Theory and Applications}, vol. 147, no.~4, pp.
  387--394, 2000.

\bibitem{Alamir:2014}
M.~Alamir, ``Fast {NMPC}: {A} reality-steered paradigm: {K}ey properties of
  fast {NMPC} algorithms,'' in \emph{European Control Conference}, Strasbourg,
  France, June 2014, pp. 2472--2477.

\bibitem{Gros:2016}
S.~Gros, M.~Zanon, R.~Quirynen, A.~Bemporad, and M.~Diehl, ``From linear to
  nonlinear {MPC}: Bridging the gap via the real-time iteration,''
  \emph{International Journal of Control}, pp. 1--19, 2016.

\bibitem{Gomes:2006}
G.~Gomes and R.~Horowitz, ``Optimal freeway ramp metering using the asymmetric
  cell transmission model,'' \emph{Transportation Research Part C: Emerging
  Technologies}, vol.~14, no.~4, pp. 244--262, 2006.

\bibitem{Munoz:2004}
L.~Mu{\~{n}}oz, X.~Sun, D.~Sun, G.~Gomes, and R.~Horowitz, ``Methodological
  calibration of the cell transmission model,'' in \emph{American Control
  Conference}, Boston, USA, July 2004, pp. 798--803.

\bibitem{Alinea:1991}
M.~Papageorgiou, H.~Hadj-Salem, and J.~Blosseville, ``{ALINEA}: {A} local
  feedback-based control law for on-ramp metering,'' \emph{Transportation
  Research Rescord}, vol.~1, no. 1320, pp. 58--67, 1991.

\bibitem{Papageorgiou:1990}
M.~Papageorgiou, ``Modeling and real-time control of traffic flow on the
  southern part of boulevard {P}eripherique in {P}aris: {Part II}:
  {C}oordinated on-ramp metering,'' \emph{Transportation Research Part A}, vol.
  24A, no.~5, pp. 361--370, 1990.

\bibitem{Hegyi:2005}
A.~Hegyi, B.~{De Schutter}, and H.~Hellendoorn, ``Model predictive control for
  optimal coordination of ramp metering and variable speed limits,''
  \emph{Transportation Research Part C}, vol.~13, no.~3, pp. 185--209, Jun.
  2005.

\bibitem{vandenWeg:2019}
G.~van~de Weg, A.~Hegyi, S.~Hoogendoorn, and B.~{D}e Schutter, ``Efficient
  freeway {MPC} by parameterization of {ALINEA} and a speed-limited area,''
  \emph{IEEE Transactions on Intelligent Transportation Systems}, vol.~20,
  no.~1, pp. 16--29, 2019.

\end{thebibliography}


\vspace{-8ex}
\begin{IEEEbiography}[{\vspace{-6ex}\includegraphics[width=1in,height=1.25in,clip,keepaspectratio]{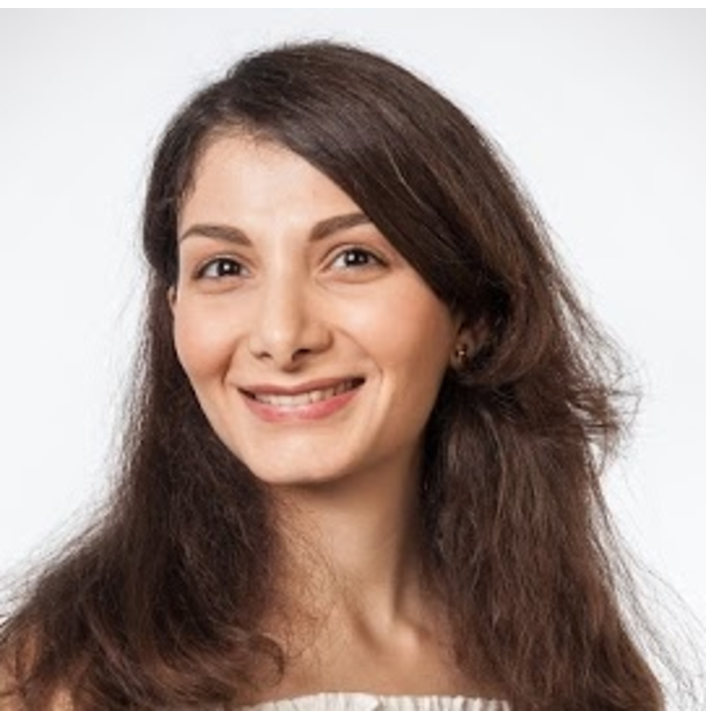}}]{Anahita Jamshidnejad}
received the PhD degree in Systems and Control from the Delft University 
of Technology, the Netherlands. 
She is currently an Assistant Professor at the Faculty of Aerospace 
Engineering, Delft University of Technology. 
Her research interests include optimization theory in engineering problems, 
integrated control methods, and model-predictive control, with applications 
in robotic systems and road traffic.
\end{IEEEbiography}
\vspace{-20ex}
\begin{IEEEbiography}[{\includegraphics[width=1in,height=1.25in,clip,keepaspectratio]{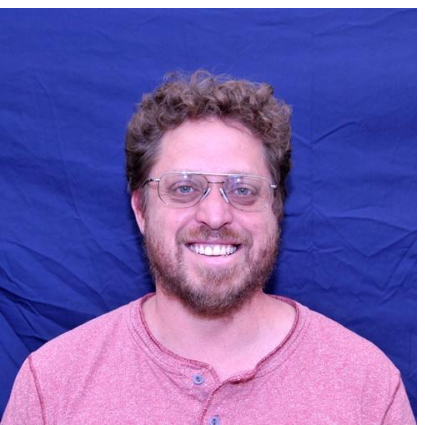}}]{Gabriel Gomes} 
received the PhD degree in systems and control theory from the 
Department of Mechanical Engineering, University of California at Berkeley. 
He is currently an Assistant Research Engineer with the Institute for Transportation Studies, 
University of California at Berkeley. 
His research focuses on the modeling, control, and simulation of transportation systems.
\end{IEEEbiography}
\vspace{-15ex}
\begin{IEEEbiography}[{\includegraphics[width=1in,height=1.25in,clip,keepaspectratio]
{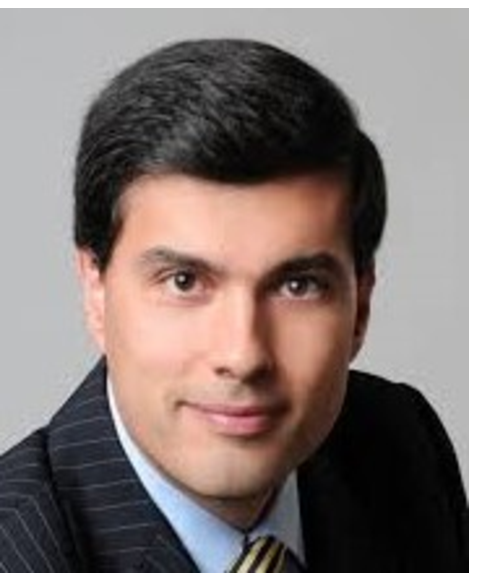}}]{Alexandre M.\ Bayen} 
is the Liao-Cho Professor of Engineering at UC Berkeley. 
He received the PhD degree in Aeronautics and Astronautics from Stanford University. 
He was a visiting researcher with the NASA Ames Research Center, from 2000 to 2003. 
Since 2014, he has been the Director of the Institute for Transportation Studies. 
He is also a Faculty Scientist in Mechanical Engineering, at the Lawrence Berkeley National 
Laboratory. 
\end{IEEEbiography}
\vspace{-15ex}
\begin{IEEEbiography}[{\includegraphics[width=1in,height=1.25in,clip,keepaspectratio]{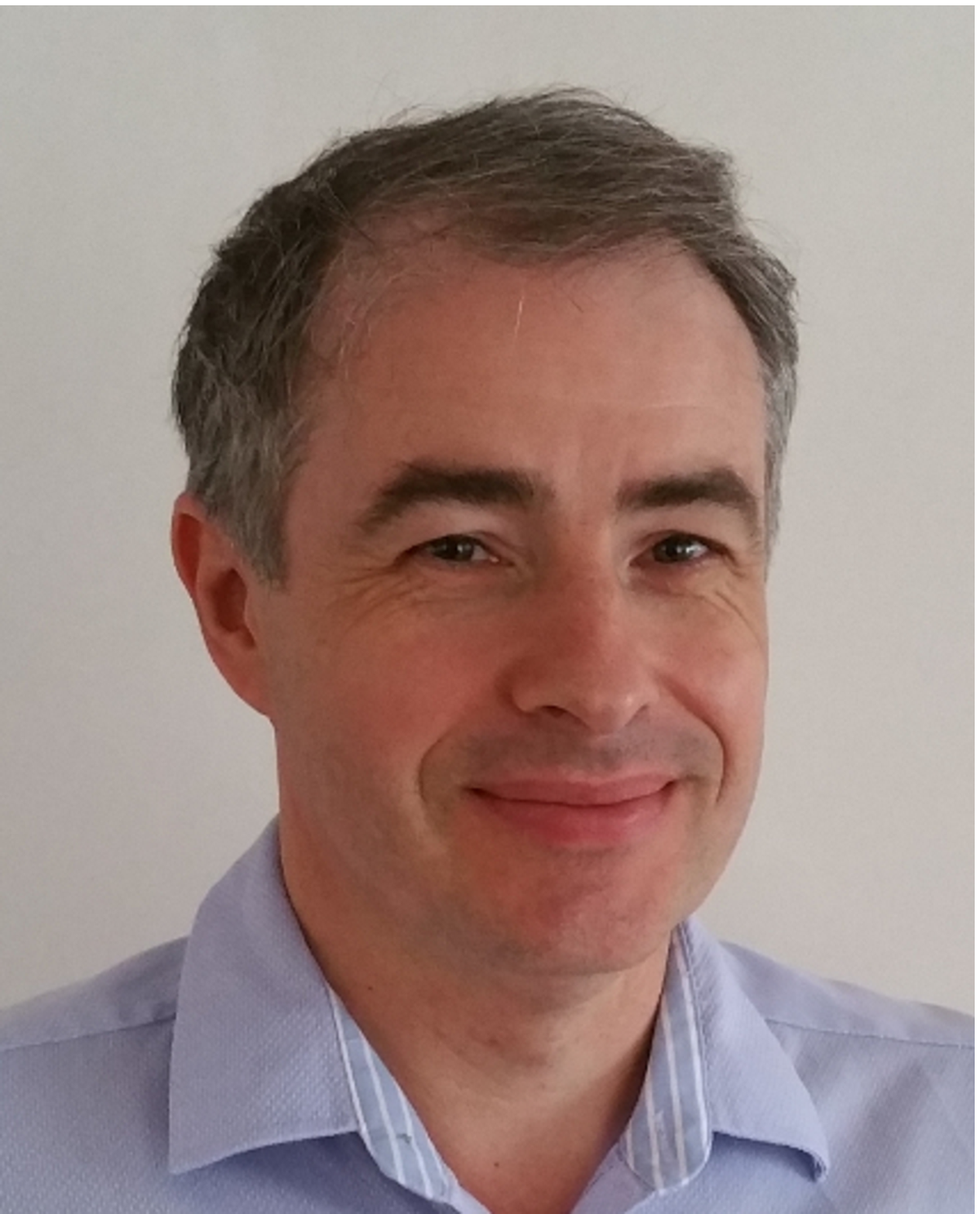}}]{Bart De Schutter }
(IEEE member since 2008, senior member since 2010, and IEEE Fellow since 2019) is
a full professor and head of department at the Delft Center for Systems and Control of Delft
University of Technology in Delft, The Netherlands. He is Senior
Editor of the IEEE Transactions on Intelligent Transportation Systems.
His current research interests include intelligent transportation and
infrastructure systems, hybrid systems, and multi-level and 
multi-agent control of large-scale systems.
\end{IEEEbiography}

\end{document}